\documentclass[12pt]{mn2e}
\usepackage{graphicx}
\usepackage{epsf}
\usepackage{lscape}
\usepackage{longtable}
\usepackage{rotating}
\usepackage{color}

\newcommand{\gtrsim}{\ga}
\newcommand{\lesssim}{\la}

\newcommand{\apj}{ApJ}
\newcommand{\apjl}{ApJ}
\newcommand{\apjs}{ApJS}
\newcommand{\mnras}{MNRAS}

\begin{document}
\topmargin -0.5in 

\title[Star-Forming Galaxies at $z\approx 8-9$ from {\em HST}/WFC3]{Star-Forming Galaxies at $z\approx 8-9$ from {\em HST}/WFC3: Implications for Reionization}

\author[Silvio Lorenzoni, et al.\ ]  
{
Silvio Lorenzoni$^{1}$\thanks{E-mail: silvio.lorenzoni@astro.ox.ac.uk}, Andrew J. Bunker$^{1}$, Stephen M. Wilkins$^{1}$,
\newauthor
Elizabeth R. Stanway$^{2}$, Matt J.\ Jarvis\,$^{3}$, Joseph Caruana$^{1}$ \\
$^1$\,University of Oxford, Department of Physics, Denys Wilkinson Building, Keble Road, OX1 3RH, U.K. \\
$^{2}$\,H.\ H.\ Wills Physics Laboratory, Tyndall Avenue, Bristol, BS8 1TL, U.K.\\
$^{3}$\,Centre for Astrophysics, Science \& Technology Research Institute, University of Hertfordshire, Hatfield, Herts AL10\,9AB, U.K.}
\maketitle

\begin{abstract}
We present a search for galaxies at $7.6<z<9.8$ using the latest {\em HST} WFC3 near-infrared data,
based on the Lyman-break technique. We search for galaxies which have large $(Y-J)$ colours (the ``$Y$-drops") on account of
the Lyman-$\alpha$ forest absorption, and with $(J-H)$ colours inconsistent with being low-redshift contaminants. 
We identify 24 candidates at redshift $z \approx 8 - 9$ (15 are robust and a further 9 more marginal but consistent with being high redshift) 
over an area of $\approx 50$ square arcminutes.
Previous searches for $Y$-drops with WFC3 have focussed only on the {\em Hubble} Ultra Deep Field (HUDF), and our larger survey (involving
two other nearby deep fields and a wider area survey) has trebelled
the number of robust $Y$-drop candidates. For the first time, we have sufficient $z\approx 8-9$ galaxies
to fit both $\phi^*$ and $M^*$ of the UV Schechter luminosity function. 
There is evidence for evolution in this luminosity function from $z=6-7$ to $z=8-9$, 
in the sense that there are fewer UV-bright galaxies at $z\approx 8-9$, consistent with an evolution mainly in $M^*$.
The candidate $z\approx 8-9$ galaxies 
we detect have insufficient ionizing
flux to reionize the Universe, and it is probable that galaxies below our detection limit provide a significant
UV contribution. The faint-end slope, $\alpha$, is not well constrained. 
However, adopting a similiar faint-end slope to that determined at $z=3-6$ ($\alpha=-1.7$)
and a Salpeter initial mass function,
then the ionizing photon budget still falls short if $f_{\mathrm{esc}}<0.5$, even integrating down to $M_{UV}=-8$.
A steeper faint end slope or a low-metallicity population (or a top-heavy IMF) might still provide sufficient
photons for star-forming galaxies to reionize the Universe, but confirmation of this might have to await
the {\em James Webb Space Telescope}.
\end{abstract} 

\begin{keywords}  
galaxies: evolution â galaxies: formation â galaxies: starburst â galaxies: high-redshift â ultraviolet: galaxies
\end{keywords} 

\section{Introduction}

The $z\approx 8$ epoch is cosmologically very interesting: the Gunn-Peterson effect (Gunn \& Peterson 1965, Scheuer 1965), the near total absorption of the continuum flux at wavelengths shorter than that of the Lyman-$\alpha$ line due to a significant  neutral hydrogen fraction in the inter galactic medium (IGM), has been observed at  $z > 6.3$ in SDSS QSO spectra
(Becker et al.\ 2001; Fan et al.\ 2001, 2006). This suggests that $z\approx 6$ lies at the end of the Epoch of Reionization, whose mid-point may have occurred at $z\approx 11$, according to latest results from WMAP (Dunkley et al. 2009).

An outstanding problem, however, is what sources were responsible for the reionization of the Universe, and when exactly this occurred.
There is evidence for old stellar populations in some $z\approx 4 - 6$ galaxies from Balmer break measurements in Spitzer/IRAC imaging (Eyles et al.\ 2005, 2007; Stark et al.\ 2007, 2009), implying star formation commenced at even earlier times. Hence it is reasonable to consider the UV photons from this star formation as a possible cause of reionization. 
Age and stellar mass determinations of the above mentioned stellar populations are affected by many uncertainties, so is important to directly look for star formation at $z > 7$ to determine whether star-forming galaxies at these epochs can indeed provide the Lyman continuum photons required for reionization.

In recent years, observations of high redshift universe ($z > 6$) have become possible. Deep imaging surveys with the {\em Hubble Space Telescope} ({\em HST}) and large ground based telescopes have made the discovery of $z\approx 6$  galaxies almost routine. Some of those searches (Bunker et al.\ 2004, Yan \& Windhorst 2004, Stanway, Bunker \& McMahon 2003, Bouwens et al.\ 2006, 2007, Oesch et al.\ 2007, Yoshida et al.\ 2006) rely on the Lyman break galaxy (LBG) technique, initially used by Steidel and collaborators (Steidel et al.\ 1996) to identify galaxies at $z\approx 3$ through the large absorption produced by the intervening Lyman-$\alpha$ forest clouds and the Lyman limit. Until recently, working at higher redshift ($z>7$) was challenging. Studies were limited to very small deep fields observed from space (e.g. Bouwens et al.\ 2008), or to extremely shallow wide-area surveys from the ground (e.g., Stanway et al.\ 2008, Hickey et al.\ 2010). The new Wide Field Camera 3 (WFC3) instrument, installed on {\em HST} in May 2009, allowed this technique to be more effectively applied to $z\approx 7 - 10$, thanks to its near-infrared channel with significantly larger field and better sensitivity than the previous-generation NICMOS instrument. Using WFC3 broad-band filters
at 1.0, 1.25 and 1.6$\mu$m (the $Y$-, $J$- and $H$-bands) and targetting fields with existing deep Advanced Camera for Surveys (ACS) data it is possible to identify optical ``drop-outs", objects seen only in the WFC3 infrared images but not in the optical ones. These are candidate $z\gtrsim 7$ galaxies. Searching for objects with no flux at 1.0$\mu$m and below ($Y$-band drop-outs, or ``$Y$-drops") could lead to the discovery of $z\approx 8$ galaxies. Deep optical images of the observed fields are still necessary to ``clean up" the list of candidates, because, as we will see later, optical detections are useful in ruling out many
lower-redshift contaminants. 

The past few months have seen several papers presenting high redshift galaxy candidates from  {\em HST}/WFC3 imaging of the {\em Hubble} Ultra Deep Field (HUDF; Bunker et al.\ 2010, McLure et al.\ 2010, Oesch et al.\ 2010, Bouwens et al.\ 2010a, Yan et al.\ 2010, Finkelstein et al.\ 2010). In the HUDF, $\sim 10$ $z'$-drops ($z\approx 7$) have been found, along with $\sim 5$ $Y$-drops ($z\approx 8$). Spectroscopic confirmation of these candidates in the
HUDF will be
extremely challenging, as they have magnitudes $J>26.5$ (for the $z'$-drops) and $J>28.0$ (for the $Y$-drops).
What is needed are larger samples over wider areas, which might yield rarer but brighter candidates more suitable for spectroscopic follow-up.
Such follow-up is important to test the validity of the $Y$-drop selection technique, and address the contaminant fraction,
as well as exploring the physics of star forming galaxies at $z\gtrsim 8$ (in particular whether Lyman-$\alpha$ emerges during the Gunn-Peterson
absorption era).
Increasing the survey area of the WFC3 LBG searches will also improve the statistics (and hence the rest-UV luminosity function constraints),
and we have started to do this by searching for $z'$-drops in the larger-area Early Release Science WFC3 images of some of the GOODS-South field (Wilkins et al.\ 2010a) and expanding this to include two other deep flanking fields (UDF-P12 \& UDF-P34) close to the HUDF (Wilkins et al.\ 2010b), which has increased the number of robust $z'$-drops from $\sim 10$ to $\sim 40$. In this paper we use our new reductions of the ERS GOODS-South, and UDF-P12 \& UDF-P34 to search for $Y$-drops at $z\approx 8 - 9$. In Bunker et al.\ (2010) we presented our preliminary list of $Y$-drops in the HUDF, and here we also re-analyse this field using a more recent data reduction.

The evolution of the rest-UV luminosity function is key to both understanding the star formation history of the Universe, and also
to address the role of star-forming galaxies in reionization. There seems to be strong evolution in the UV luminosity function (UVLF) up to redshift up to $z\approx 6$ (e.g. Stanway, Bunker \& McMahon 2003), and recent studies (Bunker et al.\ 2010; Wilkins et al.\ 2010b; Oesch et al.\ 2010) seem to show that this evolution continues up to $z\approx 7$, although based on small-number statistics. Our goal is to push the measurement of the UVLF further back in cosmic time by assembling a statistically-significant sample of probable $z\approx 8 - 9$ galaxies. From this we can address the evolution of the star formation rate density, and the ionizing photon budget.

This paper is organised as follows: in Section~2 we outline the {\em HST} observations with WFC3 and the data reduction,
and in Section~3 we describe our colour selection to recover high-redshift Lyman break galaxies, and compare our
sample with those from other studies.  In Section~4 with discuss the evolution of the star formation rate density,
and the implications for reionization, derived from the luminosity function we infer at $z\sim 8$. Our conclusions are
presented in Section~5.
Throughout, we adopt the standard concordance cosmology of $\Omega_{M}= 0.3$, $\Omega_{\Lambda}= 0.7$ and use $H_0=70$\,km\,s$^{-1}$\,Mpc$^{-1}$. All magnitudes are on the AB system (Oke \& Gunn 1983).

\section{Observations and Data Reduction}

\subsection{Observations}

In this paper we analyse images from WFC3 on {\em HST} taken in the near-infrared $Y$-, $J$- and $H$-bands.
The data come from two different {\em HST} programs, both covering areas within the GOODS-South field (Giavalisco et al.\ 2004). The {\em HST} Treasury programme GO-11563 (P.I.\ G.~Illingworth) covers the HUDF and two nearby deep flanking fields (UDF-P12 and UDF-P34,
also referred to as HUDF05-01 and HUDF05-02 in programme GO-11563).
These flanking fields were imaged by the Advance Camera for Surveys (ACS) on {\em HST}
  in $v$-, $i'$- and $z'$-bands during 2005--6 in parallel with deep {\em HST} NICMOS NIC3 observations of the original UDF as part of program GO-10632 (P.I.\ M.~Stiavelli). In Bunker et al.\ (2010) we analysed the single-WFC3-pointing HUDF data obtained soon-after the commissioning of WFC3, and in this paper we study
the two new deep WFC3 pointings on the two deep flanking fields, and reanalyse the UDF data using more recent on-orbit calibration of the detector.
Additionally, we analyse the Early Release Science (ERS) program GO/DD-11359
(P.I.\ R.~O'Connell) data, covering ten overlapping pointings with two orbits in each filter. An analysis of the first 6 pointings for $z$-drops at $z\approx 7$ was presented in Wilkins et al.\ (2010a), with the full ERS mosaic and UDF-P12\,\&\,P34 flanking fields used to select $z$-drops in Wilkins et al.\ (2010b).

The infrared
channel of WFC3 was used, which is a Teledyne $1014\times 1014$ pixel HgCdTe detector (a 10-pixel strip on the edge is not illuminated by sky and used for pedestal estimation),
with a field of view of $123"\times 136"$.
Filters used in the two programs are the same for $J$ and $H$ bands (F125W and F160W), while the ERS images use a $Y$-band filter
(F098M) which covers only the blue side of the wider F105W filter used in the UDF and flanking field images.
The data were taken in ``MULTIACCUM" mode using SPARSAMPLE100, which non-destructively reads the array every 100\,seconds. These repeated non-destructive reads of the infrared array allow gradient-fitting to obtain the count rate (``sampling up the ramp'') and the flagging and rejection of cosmic ray strikes. For the HUDF and flanking fields (Programme GO-11563) there were two exposures per orbit, with each MULTIACCUM comprising 16 reads for a total duration of 1403\,sec per exposure. For the ERS images (Programme GO/DD-11359) there were 3 exposures per orbit, each with 9 or 10 reads
and with a total time of 803--903\,s per exposure.
In Table~\ref{tab:exptimes} we list the exposure time (and number of exposures) for each field and each spectral band.
We note that for the $Y$-band images of the HUDF, two visits (8 exposures) were severely affected by image
persistence, and as in Bunker et al.\ (2010) we exclude these from our data reduction.

\begin{table}
\begin{tabular}{lccccc}
\multicolumn{6}{c}{WFC3 exposure times, in ksec (number of exposures).} \\
Field ID & $Y$-band$^{a}$ & $J$-band &$H$-band & $J$ 6\,$\sigma$ & $J$ 7\,$\sigma$\\
\hline\hline
HUDF & 39.3 (28)&44.9 (32) &78.6 (56)  & 28.51 &28.34 \\
P34  &28.1 (20)& 39.3 (28)&47.7 (34)  & 28.39 &28.22 \\
P12  &16.5 (12)&33.2 (24) &5.6 (4) & 28.31 &28.14 \\
ERS &5.0 (6)&5.0 (6) &5.0 (6) & 27.20 &27.03 \\
\end{tabular}
 $^{a}$ $Y_{098m}$ for the ERS fields and $Y_{105w}$ for the HUDF/P12/P34 fields.
 
\caption{The total exposure time (in ksec) is listed for each filter, with the number of individual
exposures given in parentheses. The final columns gives the 6\,$\sigma$ and 7\,$\sigma$ magnitude limits
in the $J$-band -- the 6\,$\sigma$ is the limit of our catalog for candidate selection,
and the luminosity function has been computed using $Y$-drops brighter than 7\,$\sigma$. All magnitudes
are on the AB system, measured in a $0\farcs6$-diameter aperture with an aperture
correction applied.}
\label{tab:exptimes}
\end{table}

\subsection{Data Reduction}

\begin{table*}
\begin{tabular}{lccccccccc}
& \multicolumn{7}{|c|}{ACS/WFC3 $2\sigma$ Detection Limits (AB magnitudes)} \\
Field ID & Center (J2000) & Area (sq.arcmin) & $b_{435w}$&$v_{606w}$&$i_{775w}$ & $z_{850lp}$ & $Y_{098m/105w}$$^{a}$ & $J_{125w}$ & $H_{160w}$ \\
\hline\hline
HUDF & 03:32:38.4 -27:47:00 & 4.2 &30.3 &30.7 &30.6 &30.0 & 29.65 &29.70 & 29.67 \\
P34  & 3:33:05.3 -27:51:23 & 4.2 & - & 29.9 & 29.5 &29.7 &29.45 &29.58 & 29.41\\
P12  & 3:33:01.9 -27:41:10 & 4.2 & - & 29.9 & 29.6 & 29.6 & 29.16 & 29.50 & 28.23 \\
ERS & 3:32:23.6 -27:42:50 & 37.0  & 29.1 & 29.1 & 28.5 & 28.4 & 28.02 & 28.39  & 28.10 \\
\end{tabular}

$^{a}$$Y_{098m}$ for the ERS field and $Y_{105w}$ for the HUDF/P12/P34 fields. 
\caption{Summary of observations. All magnitudes are on the AB system, and measured in a $0\farcs6$-diameter aperture with an aperture correction applied to correct to approximate total flux (for compact sources).}
\label{tab:obssum}
\end{table*}

The IRAF.STSDAS pipeline {\tt calwfc3} was used
to calculate the count rate and reject cosmic rays through gradient fitting,
as well as subtracting the zeroth read and flat-fielding. We used MULTIDRIZZLE
(Koekemoer et al.\ 2002) to combine exposures taken through the same filter in each pointing,
taking account of the geometric distortions and mapping on to an output
pixel size of $0\farcs06$ from an original $0\farcs13\,{\rm pix}^{-1}$.
This was the same scale as we used in our analysis of the Hubble Ultra Deep Field
WFC\,3 images (Bunker et al.\ 2010) and corresponds to a $2\times 2$ block-averaging of the
GOODSv2.0 ACS drizzled images.

The final frames had units of electrons/sec, and we take the standard ACS zeropoints for the UDF images. For WFC3,
we use the recent zeropoints reported on {\tt http://www.stsci.edu/hst/wfc3/phot\_zp\_lbn} during February 2010,
where the F098M $Y$-band has an AB magnitude zeropoint of 25.68 (such that a source of this brightness would have a count rate of 1 electron per second), and zeropoints of 26.27, 26.25 \& 25.96 for F105W, F125W \& F160W. We note that the information in
the image headers of the earlier images released in September 2009 is slightly different by 0.1-0.15\,mag, with zeropoints of F105W $Y_{ZP}=26.16$, $J_{ZP}=26.10$ \& $H_{ZP}=25.81$ (as used in Bunker et al.\ 2010).

The new WFC3 images of fields UDF-P12 and UDF-P34 were reduced using the latest version of {\tt calwfc3} (29 October 2009
release), and we also re-reduced the UDF images originally presented in Bunker et al.\ (2010) -- the
earlier paper had used a reduction using {\tt XDIMSUM} and fits to the geometric distortion due
to MULTIDRIZZLE not then being available for the newly-commissioned WFC3. For the UDF and flanking
fields, we used a pixel fraction of 0.6 to recover some of the under-sampling. In each of these three fields
we survey 4.18\,arcmin$^2$ in all exposures, with another 0.67\,arcmin$^2$ surveyed at half the maximum depth in each field.
For the ERS project, we reduced the data for all 10 pointings using the same technique as described in
Wilkins et al.\ (2010a, our analysis of the first 6 pointings). Here we used a pixel fraction of 1.0 in multidrizzle,
as we only had 6 exposures; a smaller pixel fraction would not fully populate the output repixellated grid
with such a small number of exposures (this pixel fraction was also used for the $H$-band imaging of P12, as only
4 exposures had been taken at the time of writing -- a small subset of the total still to be observed). We then mosaicked together the 10 ERS pointings in each filter, using inverse-variance weighting
for the overlap regions, producing a field of fairly uniform depth covering 37\,arcmin$^{2}$, with a further 8\,arcmin$^2$ going
less deep.

In our final combined $J$-band image, we measure a FWHM of $\approx 0\farcs1$ for point sources in the field. As most high-redshift galaxies are likely to be barely resolved (e.g., Bunker et al.\ 2004, Ferguson et al.\ 2004) we perform photometry using fixed apertures of $0\farcs6$ diameter, and introduce an aperture correction to account for the flux falling outside of the aperture. This correction was determined to be $\approx 0.2-0.25$\,mag in WFC3 from photometry with larger apertures on bright but unsaturated point sources. We note that the $H$-band images display significant Airy diffraction rings around point sources. For the ACS images, the better resolution and finer pixel sampling require a smaller aperture correction of $\approx 0.1$\,mag. All the magnitudes reported in this paper have been corrected to approximate total magnitudes (valid for compact sources), and we have also corrected for the small amount of foreground Galactic extinction
toward these fields using the {\it COBE}/DIRBE \& {\it IRAS}/ISSA dust maps
of Schlegel, Finkbeiner \& Davis (1998). The optical reddening is
$E(B-V)=0.009$, equivalent to extinctions of
$A_{850lp}=0.012$,  $A_{105w}=0.010$, $A_{125w}=0.008$ \& $A_{160w}=0.005$.

The geometric transformation and image re-gridding produces an output where the noise is highly correlated, hence measuring the standard deviation in blank areas of the final drizzled image will underestimate the noise. To ascertain the true significance of object detections, we determine
the real noise using several different techniques. As in Bunker et al.\ (2010) we also produced a crude combination of the individual flat-fielded images 
using integer-pixel shifts. While this was not used for our science (as the significant geometric distortions were not accounted
for, and it did not address the under-sampling of the PSF as ``drizzle" does), this output frame had the advantage that
the noise properties were preserved and adjacent pixels were uncorrelated. We measured the standard deviation of the counts
in blank areas of sky in this shift-and-add mosaic, and we verified that the noise (normalized per unit time) decreased as the square root of the number of frames combined. The limiting magnitudes found using these uncorrelated ``true-noise frames"  are in good agreement with the STScI {\em HST}/WFC3 Exposure Time Calculator (ETC) -- Table~\ref{tab:obssum} presents our $2\,\sigma$ limits in a $0\farcs6$-diameter aperture, with the aperture correction applied. We also produced a noise model based on the detector gain, readout noise and Poisson
counts of the measured background (including the instrument dark current), and verified that our sensitivity was well within 10 per cent of the expected noise. Finally, we measure the correlated noise (the standard deviation
of the background counts) in the drizzled image mosaics which we use for our source detection and photometry,
and use the relations in equation A13 of Casertano et al.\ (2000) to introduce a correction factor which depends on the output pixel scale 
and the size of the ``droplet" in the drizzling procedure (``pixfrac"). We generally found good agreement (at the 0.05\,mag level)
with our sensitivity measurements using the true-noise frames, except for the HUDF data where the corrected drizzle noise
underestimated the true noise by 0.1-0.2\,mag, perhaps because of the large number of frames combined with small sub-pixel shifts.
We adopted the sensitivity measurements from the true-noise frame, having checked that consistent results were produced by
the ETC, the noise model, and the noise in the drizzle frame corrected for pixel correlations. Our measured noise in the HUDF
is in good agreement with Bouwens et al.\ (2010a), but we note that McLure et al.\ (2010) appear to be $\approx 0.3$\,mag less sensitive
(although we note that their $5\,\sigma$ magnitude limit in a $0\farcs4$-diameter aperture appears not to have been corrected
to total magnitudes with an aperture correction, unlike in Bouwens et al.\ 2010a).

The optical {\em HST} ACS imaging comes from the {\em Hubble} Ultra Deep Field (Beckwith et al.\ 2006), and we used
the publicly-available $v,i,z$ reductions of flanking field UDF-P12 provided by the UDF05 team (Oesch et al.\ 2007).
We reduced the $v,i,z$ ACS data for UDF-P34 from the {\em HST} archive, using MULTIDRIZZLE to combine a large subset
of the data comprising blocks of data taken
close in time with similar telescope roll angles, again using an output $0\farcs03$ pixel scale. These subsets of drizzled images were then registered and
combined with {\tt IRAF.imcombine}. Our combined images were 4.8\,ksec in $v$-band, 10.6\,ksec in $i$-band and 26.8\,ksec in $z$-band.
All the ACS images were then block-averaged $2\times 2$ and registered with our drizzled WFC3 frames.

\subsection{Construction of Catalogues}

To perform the candidate selection we used the SExtractor photometry package (Bertin
\& Arnouts 1996), version 2.5.0. Since we are searching for $Y$-drops (objects clearly detected in the WFC3 $J$-band but with minimal flux in the $Y$-band and ACS images), fixed circular apertures $0\farcs6$ in diameter were `trained' in the $J$-image, and running SExtractor in dual-image mode those apertures were used to measure the flux in the same locations in the $Y$-band image.
The same procedure was repeated between $J$-band image and all the other ACS and WFC3 images with different filters. For each waveband we used a weight image derived from the exposure map.
To identify the objects we set the SExtractor parameters to have a lower limit of 5 contiguous pixels above a threshold of $2\sigma$ per pixel (data were drizzled to a
scale of 0\farcs06~pixel$^{-1}$). We corrected the aperture magnitudes to approximate total
magnitudes with the aperture correction appropriate for that filter. With this cut we were able to detect all significant sources, along with some spurious detections just above the noise limit or due to diffraction spikes from stars.
We also impose a 6\,$\sigma$ limit on the $J$-band magnitude for all fields, with the $J_{AB}$ magnitude limit listed in the last column of Table \ref{tab:exptimes}.
Table~\ref{tab:objects} presents our photometry of $Y$-drops from SExtractor, where we have corrected the magnitude errors returned by SExtractor
for the effects of correlated noise in the drizzled images, using our ``true noise frames" to determine the scaling factor (typically SExtractor
underestimated the magnitude errors by a factor of $\approx1.5$ for pixfrac=0.6 used in most of our data, and a factor of $\approx2.6$ for pixfrac=1.0 as used in the ERS
and the $H$-band of P12).

\section{Candidate Selection}

Identification of candidates is achieved using the Lyman break
technique (e.g. Steidel et al. 1996), where a large colour
decrement is observed between filters either side of
 Lyman-$\alpha$  in the rest-frame of the galaxy. At $z>6$, the flux
 decrement comes principally from the large 
integrated optical depth of the intervening absorbers
(the Lyman-$\alpha$ forest).

At $z\approx 8-9$ the location of the Lyman-$\alpha$ break is redshifted to $\sim 1.1\mu m$ -- the WFC3 $Y_{105w/098m}$ and $J_{125w}$ are suitably located such that a $7.6<z<9.8$ star forming galaxy will experience a significant flux decrement between these two filters (see Figure \ref{fig:lbg_spectra} \& Figure~\ref{fig:selection}), although the selection efficiency drops at the extremes of this range. 

\subsection{Contamination}

Searching for distant galaxies using only broadband photometry means that contamination is a potentially serious issue. There are two main sources of contamination: objects whose intrinsic colours are similar to those of the target population; and faint objects with intrinsically different colours but whose observed colours scatter into our selection because of photometric noise. We note that the effect of transient phenomena is not significant for the selection of $Y$-drops, since the WFC3 $Y$, $J$ \& $H$ images were taken close in time. This is unlike our selection of $z'$-drops (e.g. Wilkins et al.\ 2010a, 2010b; Bunker et al.\ 2010) where the ACS $z'$-band and WFC3 $Y$-band were separated by many years, so a transient such as a supernova or high-proper-motion object which entered the $Y$-band but was absent at that location in the ACS could be erroneously identified as a Lyman-break galaxy. Indeed, a probable supernova was identified in the WFC3 imaging of the HUDF (e.g., Bunker et al.\ 2010).

\subsubsection{Intrinsically Red Objects}

\begin{figure}
\centering
\includegraphics[width=18pc]{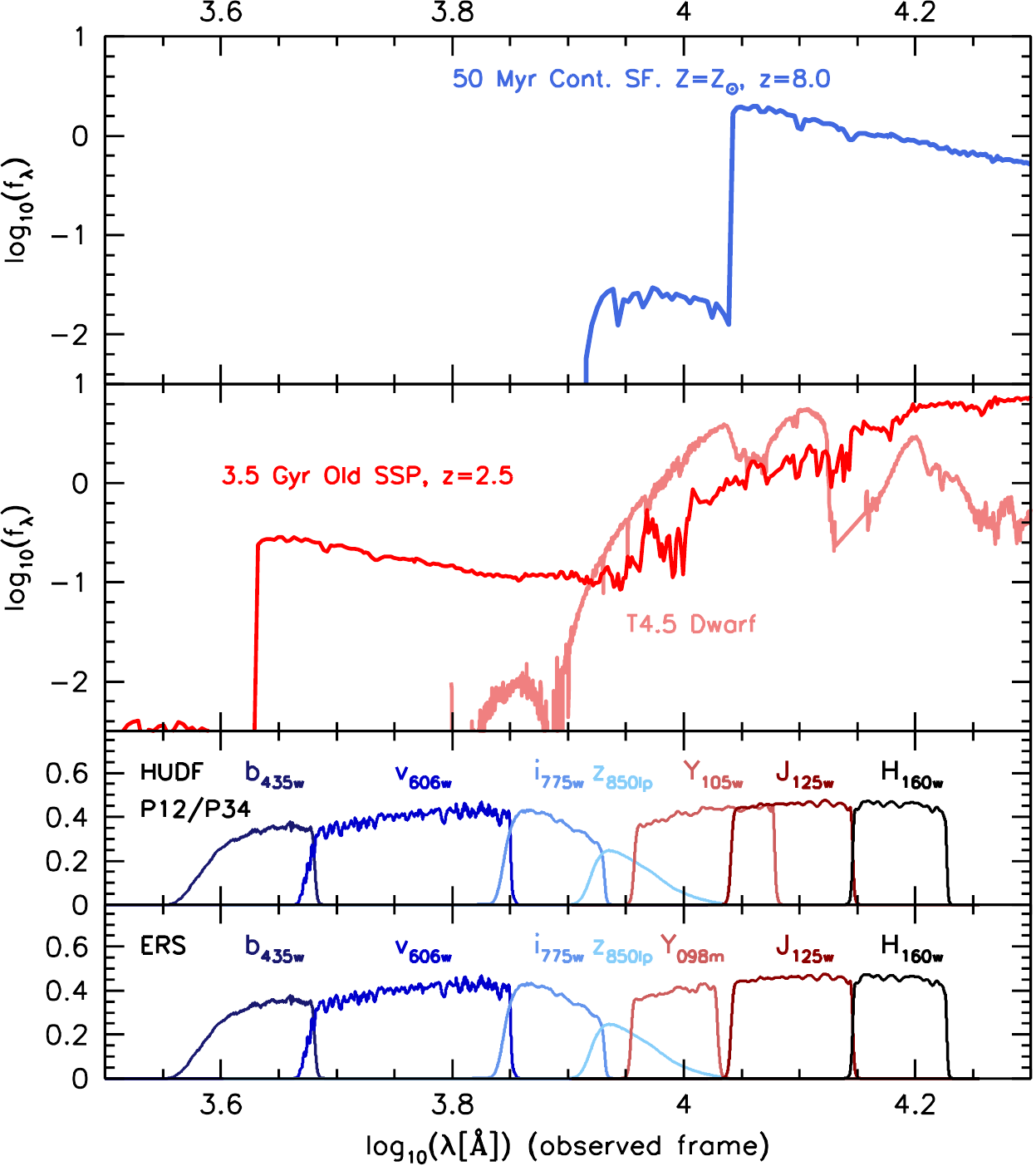}
\caption{Top panel - Model (from the Starburst99, Leitherer et al. 1999) spectral energy distribution (SED) of a redshifted $z=8$ star forming galaxy. Middle panel - Potential contaminants: Observed SED of a low-mass dwarf star (class: $T4.5$, Knapp et al. 2004) together with the model (Starburst99) SED of a $3.5$Gyr Single-aged Stellar Population (SSP) at $z=2.5$. The bottom two panels show the transmission functions of the combination of filters available to each field.}
\label{fig:lbg_spectra}
\end{figure}

There are two distinct types of objects whose apparent $Y_{105w/098m}-J_{125w}$ colours are similar to those of Lyman-break galaxies at $z\approx 8$: lower-redshift ($z\approx 2$) galaxies have the Balmer/$4000{\rm \AA}$ break feature between the two filters used, $Y_{105w/098m}$ and $J_{125w}$, while some low mass dwarf stars, especially those of L and T spectral class, have low temperatures and broad absorption features that can mimic a spectral break.

\begin{figure}
\centering
\includegraphics[width=16pc]{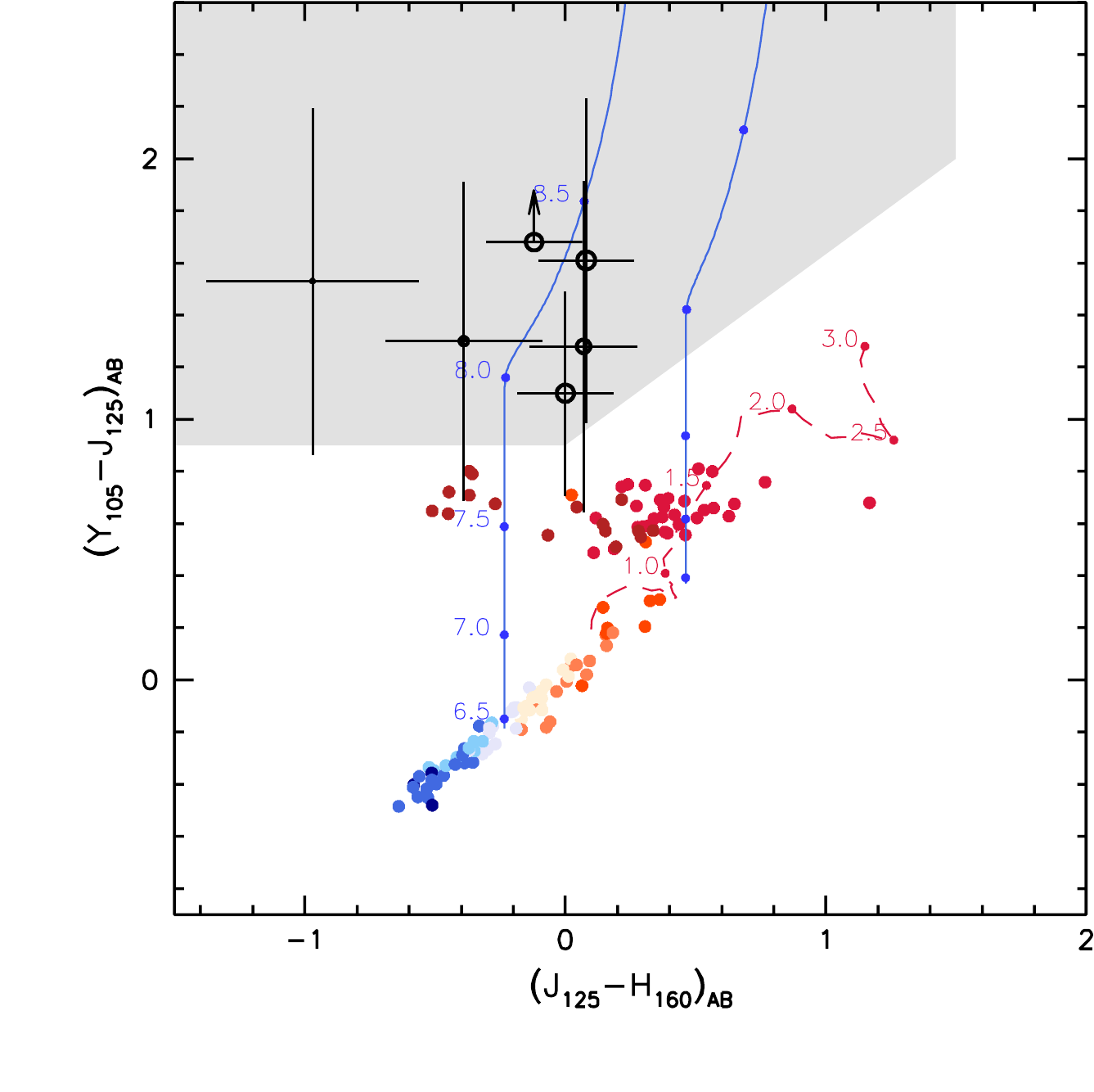}
\includegraphics[width=16pc]{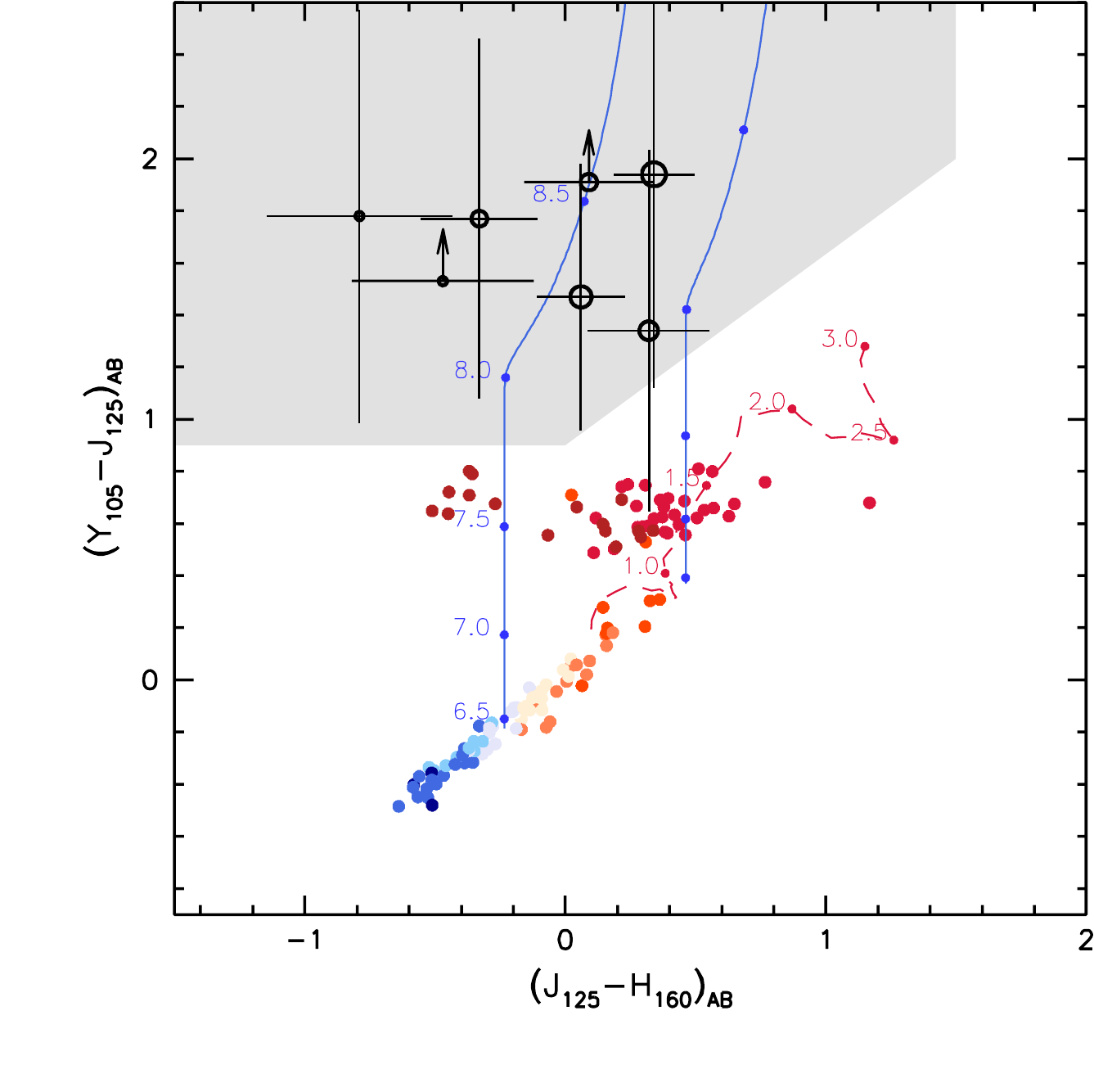}
\includegraphics[width=16pc]{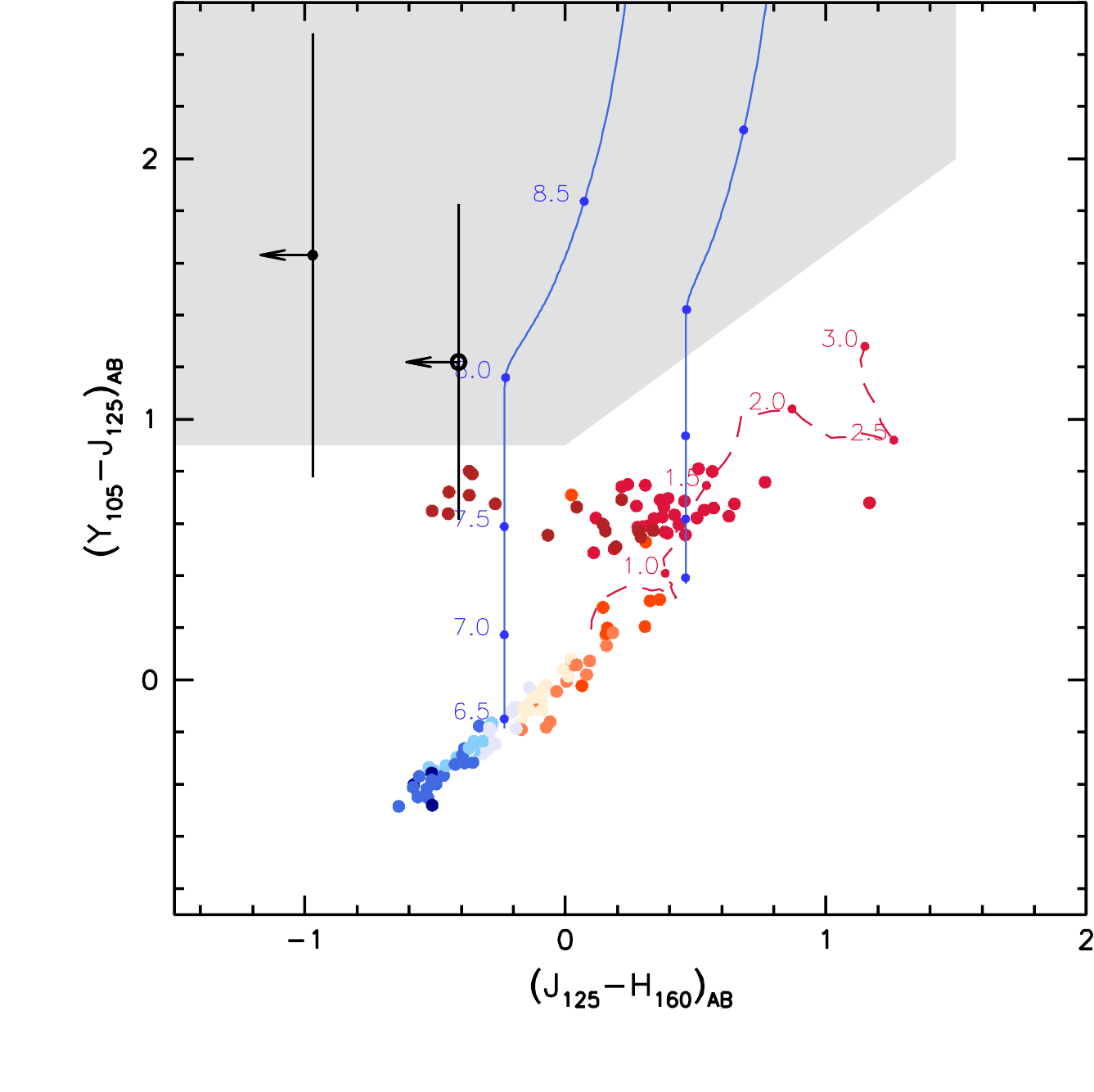}
\caption{$J_{125w}-H_{160w}$ and $Y_{105w}-J_{125w}$ colour - colour figures for the HUDF (top) and flanking fields P34 (middle) and P12 (bottom), showing our $YJH$ colour selection window (grey shaded area), the location of our candidates, the predicted paths taken by high-redshift galaxies (solid lines, $\beta=-3.0$ , left, and $\beta=0.0$, right) and the location of possible contaminating sources. Contaminating sources include Galactic stars (O - T dwarf stars, with L and T stars being redder, denoted by filled circles) and a passively evolving `early-type' galaxy (modelled as an instantaneous burst of star formation at $z=10$ followed by passive luminosity evolution, denoted by the dashed line). High-redshift candidates are denoted by black circles (where the size of the circle is an indication of the apparent $J_{AB}$ magnitude,
with bigger circles indicating brighter sources). Limits and error bars are  $1\,\sigma$. 
}
\label{fig:cc_1}
\end{figure}

\begin{figure}
\centering
\includegraphics[width=16pc]{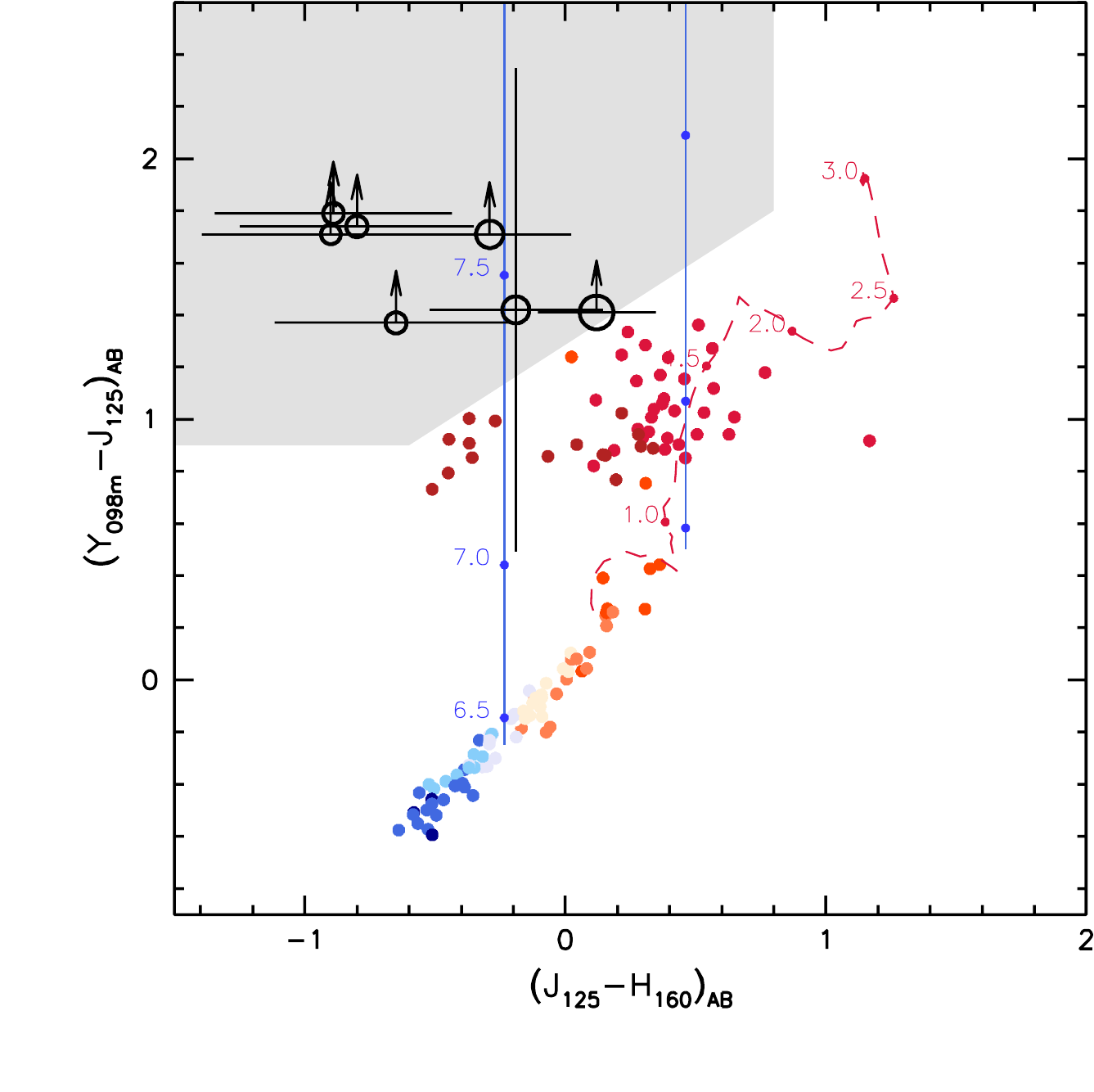}
\caption{$J_{125w}-H_{160w}$ and $Y_{098m}-J_{125w}$ colour - colour figures for the ERS field showing our $YJH$ colour selection window (grey shaded area), the location of our candidates, the predicted paths taken by high-redshift galaxies (solid lines, $\beta=-3.0$ , left, and $\beta=0.0$ , right) and the location of possible contaminating sources. Contaminating sources include galactic stars (denoted by filled circles) and a passively evolving instantaneous burst of star formation that occured at $z=10$ (dashed line). High-$z$ candidates are denoted by black circles (where the size of the circle is an indication of the apparent $J_{AB}$ magnitude, with the brighter sources being bigger circles). Limits and errorbars are $1\,\sigma$. Objects ERS.YD7 and ERS.YD8 are very marginal and hence not 
included in this figure (see Section 3.2).}
\label{fig:cc_2}
\end{figure}

Examples of the spectral energy distributions (SEDs) of each of these types of object (a model $3.5$ Gyr old single-aged stellar population at $z=2.5$ and a T4.5 dwarf star) are shown in Figure \ref{fig:lbg_spectra}. In the case of lower redshift galaxies the slope of the SED longward of the spectral break (i.e. longward of $Y_{105w/098m}$) is somewhat redder than that predicted for a high-$z$ star forming galaxy. The addition of a further filter at wavelengths redder than the $J_{125w}$ filter ($H_{160w}$ in this case) can then be used to discriminate between high-$z$ and lower redshift galaxies (Figure \ref{fig:cc_1}). L and T dwarfs contamination in the HUDF and P34 field is mostly ruled out by the $Y_{105w}-J_{125w}$ colour selection we adopted. The addition of $H_{160w}$ photometry is still important in excluding these objects in the ERS field (see Figure \ref{fig:cc_2}), where the different $Y$-band filter used provides less good discrimination using $Y-J$ colour alone.

In Figures \ref{fig:cc_1} and \ref{fig:cc_2} the positions of both the interlopers and the tracks expected for high-redshift star forming galaxies are shown in the $(J_{125w} - H_{160w})$ - $(Y_{105w/098m} - J_{125w})$ colour plane. With the exception of the lowest temperature T dwarfs where the $Y_{098m}$ filter is employed (the ERS field), these interlopers form a distinct locus separate from $z\approx 8-9$ star forming galaxies with UV spectral slope index $\beta<0.0$ (where $f_{\lambda}=\lambda^{\beta}$ is used as a model of the UV properties of star forming galaxies). 

Using this analysis it is possible to design a window in $(Y_{105w/098m} - J_{125w})$ -  $(J_{125w} - H_{160w})$ colour - colour space that selects mainly high-redshift star forming galaxies, while eliminating known contaminant populations. For the HUDF/P12/P34 fields (i.e. where we have $Y_{105w}$ imaging) this $YJH$ selection selection criteria is:
\begin{eqnarray*}
(Y_{105w}-J_{125w})& > & 0.9\\
(Y_{105w}-J_{125w})& > & 0.73\times (J_{125w}-H_{160w})+0.9\\
(J_{125w}-H_{160w}) & < & 1.5\\
\end{eqnarray*}

The use of an alternative $Y$ filter ($Y_{098m}$) in the ERS field necessitates the use of a slightly different criteria:
\begin{eqnarray*} 
(Y_{098m}-J_{125w})& > & 0.9\\
(Y_{098m}-J_{125w})& > & 0.64\times (J_{125w}-H_{160w})+1.28\\
(J_{125w}-H_{160w}) & < & 0.8\\
\end{eqnarray*} 

We have designed our selection criteria to reject all known interlopers, while selecting most $z\approx 8-9$ star-forming galaxies.
Other groups have used similar colour:colour selection, but with slightly different colour cuts (e.g., Bouwens et al.\ 2010a).
Although this may affect the surface density of candidates (due to a slightly different redshift range and spectral range of spectral slopes
probed for the LBGs, and a different contaminant fraction), the inferred luminosity functions should be similar as these
selection effects are corrected for in the effective volume calculation.
The window we obtain with such criteria excludes a hypothetical population of $z\approx 8-9$ galaxies with $J_{125w}-H_{160w} \gtrsim 1$ colours. Such a population would have extremely red UV spectral slopes, possibly due to massive dust reddening. 
The potential bias of our selection criteria and a more general analysis of the UV properties of the candidates presented in this work is discussed in more detail in Wilkins et al.\ (2010c) where we conclude that the distribution of UV spectral slope indices is consistent with being blue, with $\beta\approx-2$.

\subsubsection{Photometric Scatter}

Even with the selection criteria described above, we cannot prevent some objects being scattered into our selection window because of photometric noise. At low signal to noise ratio, this contamination could be significant, and we impose another criterion to exclude those objects. 
To do that we use the deep optical imaging available in the ACS $b_{435w}$, $v_{606w}$, $i_{775w}$ and $z_{850lp}$ bands: because of the strong Lyman-$\alpha$ forest absorption, $z\approx 8-9$ galaxies should not have any significant flux in the $b_{435w}$, $v_{606w}$ and $i_{775w}$ bands, so we impose an additional $bvi$ {\em non-detection} criteria for the selection of our candidates. All objects with a $>2\sigma$ detection in any of the $b_{435w}$, $v_{606w}$ and $i_{775lp}$ are classified as contaminants. The depths of these ACS images are given in Table~2. We note
that the $z_{850w}$ filter does have a red tail which overlaps with the $Y_{098m/105w}$-band filters used, so it is conceivable that
a $Y$-drop towards the lower end of the redshift selection might have residual $z$-band flux. However, we found only one $Y$-drop
(P34.YD5) with a $\sim 2\,\sigma$ detection in $z$-band.

\subsection{Candidate Galaxies at $z\approx 8-9$}

\begin{figure}
\centering
\includegraphics[width=16pc]{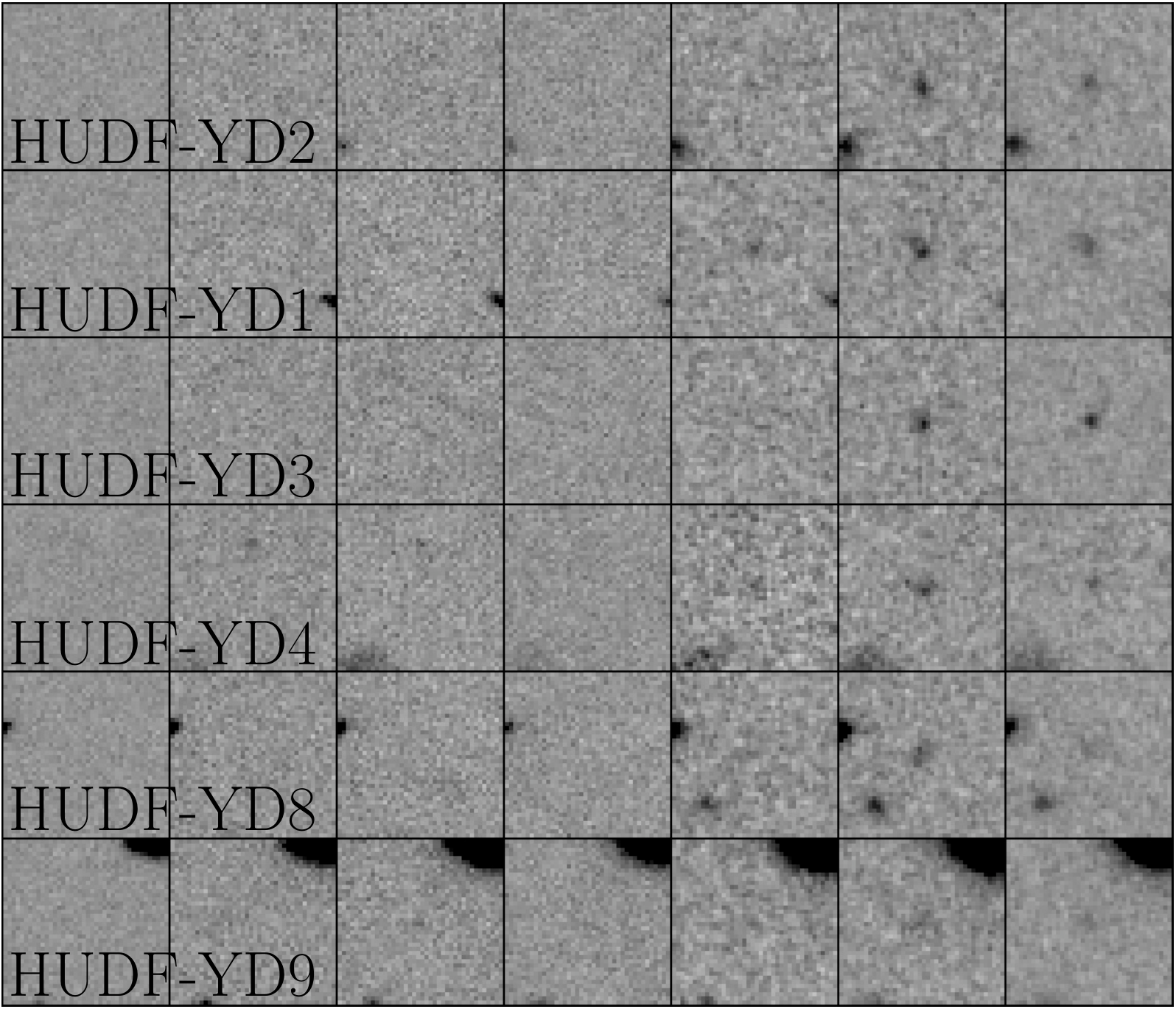}
\includegraphics[width=16pc]{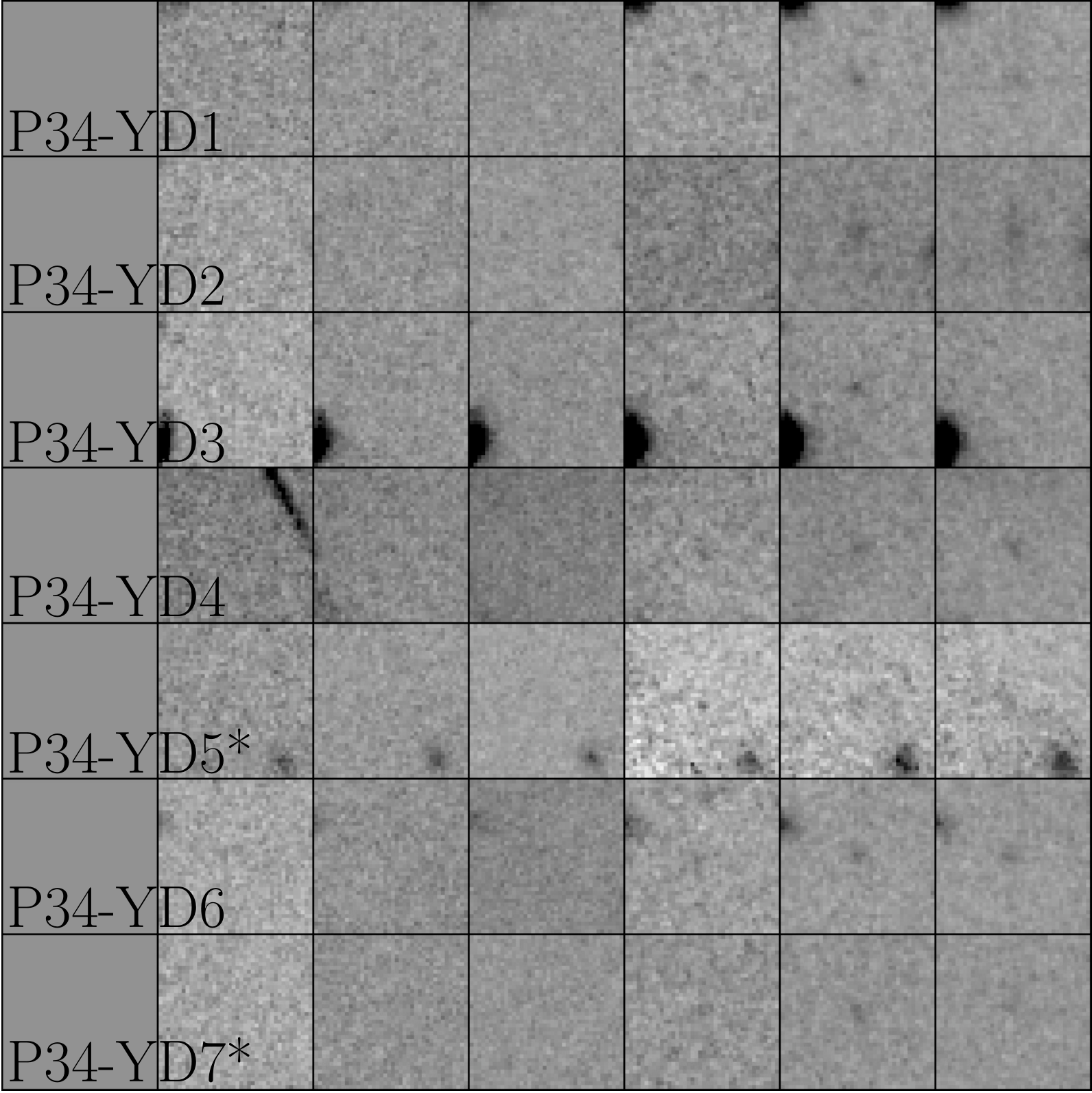}
\includegraphics[width=16pc]{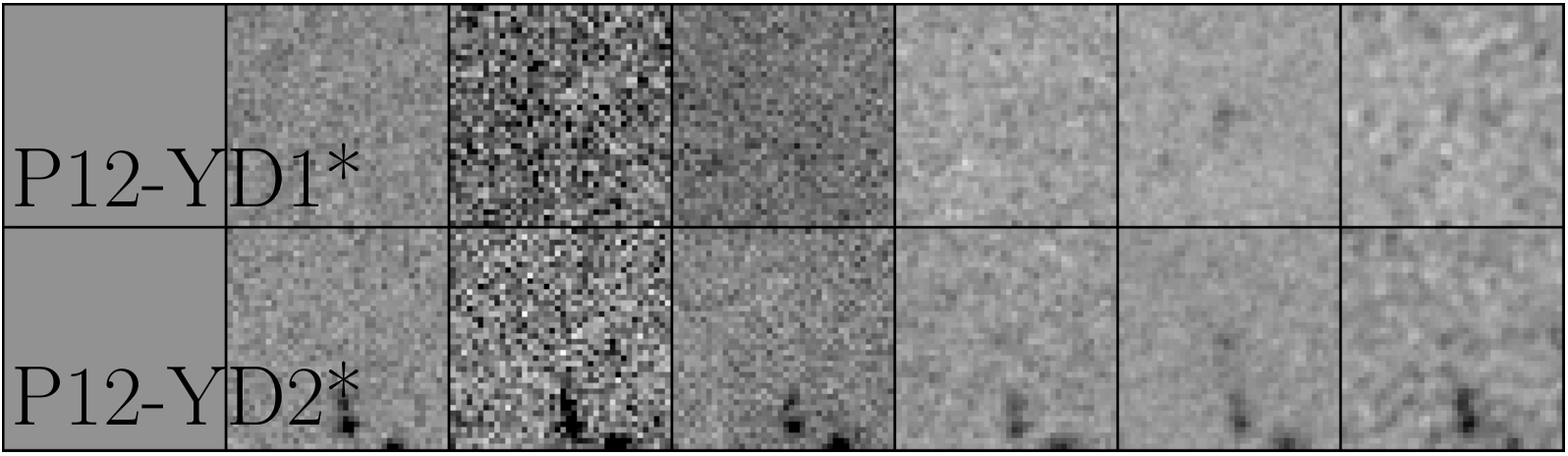}
\includegraphics[width=16pc]{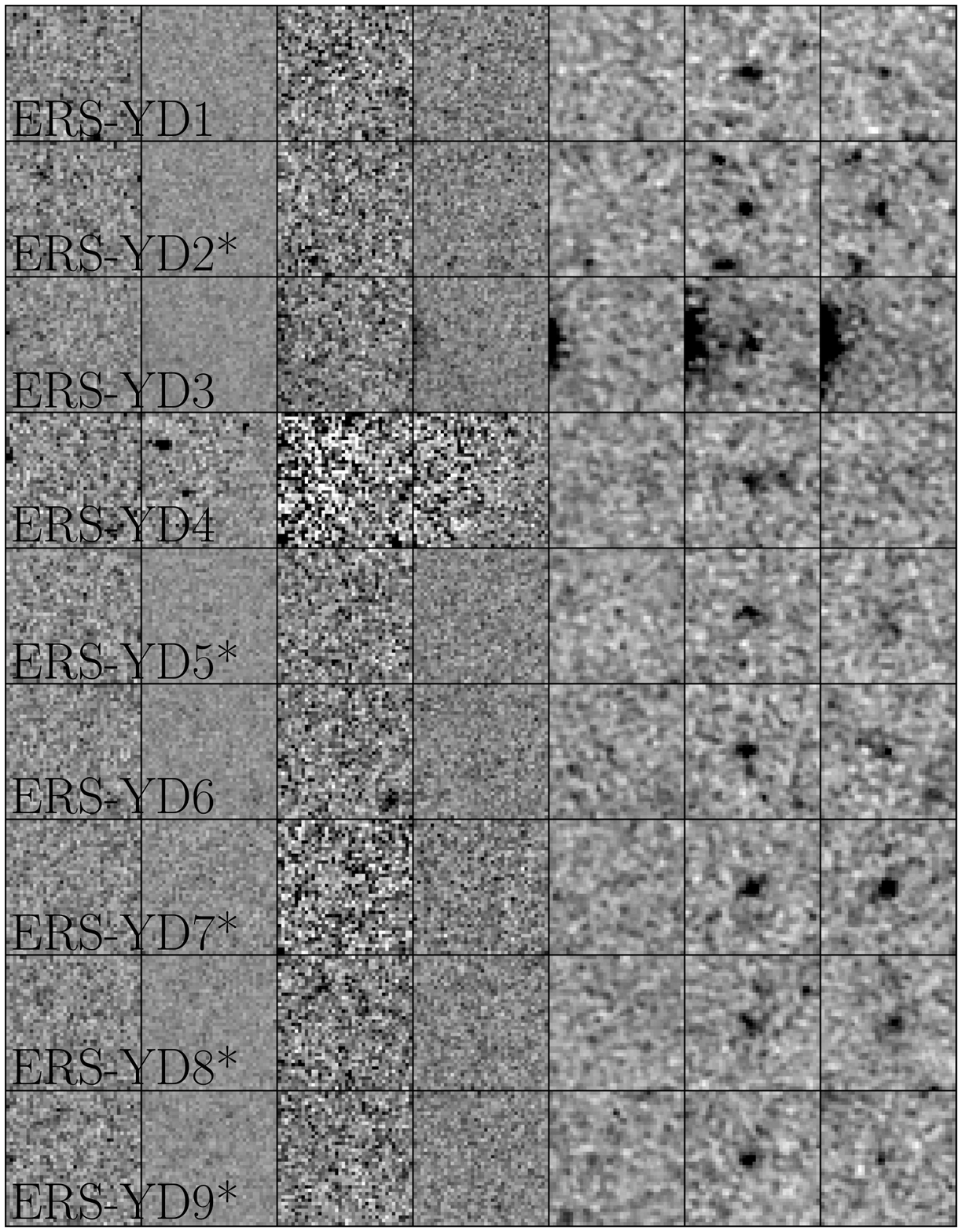}
\caption{$2\farcs4\times 2\farcs4$ $(b)vizYJH$ thumbnail images of objects meeting our selection criteria in the analyzed fields. Within
each field they are ordered by $H$-band magnitude (brightest at the top). Objects marked with * are not more marginal candidates, see Section 3.2. The fields UDF-P12 and UDF-P34 do not have ACS $b$-band imaging.}
\label{fig:stamps}
\end{figure}

After imposing our selection criteria we are able to compile a list of candidate $z\approx 8-9$ star forming galaxies in the HUDF, UDF-P34, UDF-P12 and ERS fields. In Table \ref{tab:objects} we list positions and photometry of these objects, while thumbnails of the $bvizYJH$ images of these candidates (where available) are presented in Figure \ref{fig:stamps}. In total we find 24 $Y$-drop candidates (HUDF:6, UDF-P34:7, UDF-P12:2, ERS:9) covering a range of apparent $J_{AB}$ magnitudes of $27.0-28.5$.
In the three deep single WFC3 pointings, the number of candidates is fairly consistent from field to field, with 3, 4 and 2 $Y$-drops for the HUDF, UDF-P34 and UDF-P12  fields, respectively, at $J_{AB}<28.2$.

There are 9 (of the 24) objects in the $Y$-drop list (Table~\ref{tab:objects}) which we flag as being more marginal than
the other candidates as they sit at the limits of our selection, although they are plausible $z\approx 8-9$ galaxies (our effective volume calculation
already corrects for galaxies excluded as lying just outside the selection region).
Candidates ERS.YD7 and ERS.YD8 in the ERS are flagged, as we only
have a lower limit on the $(Y-J)$ colour (they are $\lesssim\,1\,\sigma$ in $Y$-band). Adopting the $1\,\sigma$ lower limit on the $(Y-J)$ colour places them in or above the `contaminant' triangular region of Figure \ref{fig:cc_1}, fully consistent with entering our selection area. Similarly, objects ERS.YD2, ERS.YD5, ERS.YD9 and P34.YD7 are flagged: using a $1\,\sigma$ lower limit on the $(Y-J)$ colour these candidates would fully meet our selection criteria (see Figures \ref{fig:cc_1} and \ref{fig:cc_2}), while a more conservative $2\,\sigma$ lower limit could potentially locate them just below our selection box, although with colours consistent with falling within the selection window. Deeper $Y$-band imaging is required to show unambiguously that they are not in the `contaminant' region of the colour:colour space.
Object P34.YD5 in P34 is also flagged, because it has a $\sim$\,2\,$\sigma$ detection in the $z$-band. There are no detections in $v$-, $i$- and $Y$-bands, though, so it is still a likely high-redshift ($z>6$) object -- the $z$-band flux might be
statistical fluctuation or perhaps a high-equivalent-width emission line within the $z$-band.

We also flag as marginal two potential high redshift galaxies in field UDF-P12, on
the grounds that the short exposure time of the $H$-band image in this field (Table \ref{tab:exptimes}) made it impossible to measure the $H_{AB}$  magnitude. The upper limits on the $(J-H)$ colours place them away from the red contaminant region with $(J-H)>1.5$ (Figure~\ref{fig:cc_1}), but we require the UV luminosity in the $H$-filter (uncontaminated by the effects of Lyman-$\alpha$ forest absorption) to infer the absolute UV magnitude (as described in Section~\ref{sec:lumfunc}). We now consider whether these single-band detections might be due to transients (such as was the case for the likely supernova in the WFC3 images of the HUDF, object zD0 in Bunker et al.\ 2010).
The P12 field was observed in $J$-band in two observing blocks, with 8 frames taken on U.T.\ 2009 November 02, and the other 16 frames taken over U.T.\  2009 November 10--15. As a check, we combined the two different epochs separately with ``multidrizzle".
The magnitude of P12.YD1 is consistent between the two epochs, with $J=28.07\pm 0.25$ ($4.3\,\sigma$) and $J=27.95\pm 0.16$ ($6.8\,\sigma$) respectively. However, P12.YD2 might show some variability in the $J$-band with $J=27.36\pm 0.13$ ($8.3\,sigma$) for the first block of data and $J=28.14\pm 0.19$ ($5.8\,sigma$) for the second.
Hence it is plausible that P12.YD2 might be a transient rather than a high-redshift $Y$-drop.
When this WFC3 program (GO-11563) is complete, the $H$-band will be much deeper on P12, allowing a further check on the robustness
of the candidates in this field.
However, the two candidates in P12 represent less than 10 per cent of our $Y$-drop sample, so will not quantitively affect our conclusions; for the moment, we exclude this field from our fitting of the UV luminosity function.

\begin{table*}
\begin{tabular}{lccccccc}
ID & RA & Dec & $Y_{\rm AB}$ & $J_{\rm AB}$ (significance, $\sigma$) & $H_{\rm AB}$ & $(Y-J)_{AB}$ & $(J-H)_{AB}$ \\
\hline\hline
HUDF.YD2 & 03:32:37.796  & -27:46:00.12 & $ > 29.76$ & $28.08 \pm 0.12$ ($8.9\sigma$) & $28.20 \pm 0.14$ & $>1.66$ & $-0.12$ \\
HUDF.YD1 & 03:32:42.873  & -27:46:34.58 & $29.25 \pm 0.37$ & $28.15 \pm 0.13$ ($8.3\sigma$) & $28.15 \pm 0.13$ & $1.10$ & $0.00$ \\
HUDF.YD3 & 03:32:38.135  & -27:45:54.03 & $29.79 \pm 0.61$ & $28.18 \pm 0.13$ ($8.1\sigma$) & $28.10 \pm 0.13$ & $1.61$ & $0.08$ \\
HUDF.YD4 & 03:32:33.130  & -27:46:54.49 & $29.85 \pm 0.65$ & $28.32 \pm 0.15$ ($7.1\sigma$) & $29.29 \pm 0.38$ & $1.53$ & $-0.97$ \\
HUDF.YD8 & 03:32:43.082  & -27:46:27.75 & $29.75 \pm 0.59$ & $28.45 \pm 0.17$ ($6.3\sigma$) & $28.84 \pm 0.25$ & $1.30$ & $-0.39$ \\
HUDF.YD9 & 03:32:36.360  & -27:46:23.35 & $29.78 \pm 0.61$ & $28.50 \pm 0.18$ ($6.1\sigma$) & $28.43 \pm 0.17$ & $1.28$ & $0.07$ \\
\hline
P34.YD1 & 03:33:00.340  & -27:51:20.97 & $29.71 \pm 0.68 $ & $27.94 \pm 0.12$ ($9.1\sigma$) & $28.27 \pm 0.19$ & $1.77$ & $-0.33$ \\
P34.YD2 & 03:33:09.657  & -27:51:16.45 & $29.89 \pm 0.81$ & $27.95 \pm 0.12$ ($9.0\sigma$) & $27.61 \pm 0.10$ & $1.94$ & $0.34$ \\
P34.YD3 & 03:33:07.474  & -27:51:14.89 & $29.85 \pm 0.78$ & $28.07 \pm 0.14$ ($8.1\sigma$) & $28.86 \pm 0.33$ & $1.78$ & $-0.79$ \\
P34.YD4 & 03:33:04.857  & -27:51:38.28 & $29.36 \pm 0.50$ & $27.89 \pm 0.11$ ($9.5\sigma$) & $27.83 \pm 0.13$ & $1.47$ & $0.06$ \\
P34.YD5* & 03:33:06.596  & -27:52:48.95 & $ > 30.20$ & $28.29 \pm 0.17$ ($6.6\sigma$) & $28.20 \pm 0.18$ & $>1.91$ & $0.09$ \\
P34.YD6 & 03:33:06.901 & -27:51:32.54 & $29.69 \pm 0.67 $ & $28.35 \pm 0.18$ ($6.2\sigma$) & $28.03 \pm 0.15$ & $1.34$ & $0.32$ \\
P34.YD7* & 03:33:09.286 & -27:51:32.22 & $ > 29.91$ & $28.38 \pm 0.18$ ($6.1\sigma$) & $28.85 \pm 0.32$ & $>1.53$ & $-0.47$ \\
\hline
P12.YD1* & 03:33:03.034  & -27:41:57.00 & $29.64 \pm 0.84$ & $28.01 \pm 0.14$ ($7.9\sigma$) & $> 28.98$ & $1.63$ & $<-0.97$ \\
P12.YD2* & 03:33:00.545  & -27:41:46.80 & $29.25 \pm 0.59$ & $28.03 \pm 0.14$ ($7.8\sigma$) & $> 28.44$ & $1.22$ & $<-0.41$ \\
\hline
ERS.YD1 & 03:32:23.369  & -27:43:26.53 & $ > 28.77$ & $26.98 \pm 0.15$ ($7.3\sigma$) & $27.87 \pm 0.43$ & $> 1.79$ & $-0.89$ \\
ERS.YD2* & 03:32:02.986  & -27:43:51.95 & $ > 28.39$ & $26.98 \pm 0.15$ ($7.3\sigma$) & $26.86 \pm 0.17$ & $> 1.41$ & $0.12$ \\
ERS.YD3 & 03:32:29.790  & -27:43:01.09 & $ > 28.77$ & $27.03 \pm 0.16$ ($7.0\sigma$) & $27.83 \pm 0.42$ & $> 1.74$ & $-0.80$ \\
ERS.YD4 & 03:32:40.904  & -27:40:12.37 & $ > 28.77$ & $27.06 \pm 0.16$ ($6.8\sigma$) & $27.96 \pm 0.47$ & $> 1.71$ & $-0.90$ \\
ERS.YD5* & 03:32:18.414  & -27:43:45.99 & $ > 28.77$ & $27.06 \pm 0.16$ ($6.8\sigma$) & $27.35 \pm 0.27$ & $> 1.71$ & $-0.29$ \\
ERS.YD6 & 03:32:05.022  & -27:45:53.93 & $28.61 \pm 0.91$ & $27.19 \pm 0.18$ ($6.1\sigma$) & $27.38 \pm 0.28$ & $1.42$ & $-0.19$ \\
ERS.YD7* & 03:32:41.676  & -27:41:27.48 & $ > 28.03$ & $26.95 \pm 0.14$ ($7.6\sigma$) & $26.56 \pm 0.13$ & $> 1.08$ & $0.39$ \\
ERS.YD8* & 03:32:37.927  & -27:42:20.78 & $ > 28.14$ & $27.14 \pm 0.17$ ($6.3\sigma$) & $26.89 \pm 0.18$ & $> 1.00$ & $0.25$ \\
ERS.YD9* & 03:32:27.014 & -27:44:31.29 & $ > 28.57$ & $27.20 \pm 0.18$ ($6.0\sigma$) & $27.85 \pm 0.43$ & $> 1.37$ & $-0.65$ \\
\hline
\end{tabular}
\caption{$Y$-band drop out candidate $z\approx 8-9$ galaxies meeting our selection criteria in the $H$UDF, P12, P34 and ERS fields. Objects are divided by field and then ordered by apparent $J_{AB}$ magnitude. Where the measured flux is $<1\,\sigma$, an upper limit at the $1\,\sigma$
level is quoted. The significance of the $J$-band detection is shown in parentheses after the magnitude. Objects marked with * are more marginal candidates, see Section 3.2. The ERS field uses the narrower $Y$-band filter F098M, and the other fields use F105W.
For the P34 field, we introduce a small offset to the astrometry in the HST image headers (given in Lorenzoni et al.\
http://arxiv.org/abs/1006.3545v1) of $\Delta RA=0\fs03$ to match the astrometry from GSC-2 and 2MASS.}
\label{tab:objects}
\end{table*}

\subsection{Comparison with Other Studies}

We now compare our new list of candidates within the HUDF field with other groups' previous studies (Oesch et al.\ 2010, Bouwens et al.\ 2010a, McLure et al.\ 2010, Yan et al.\ 2010 and Finkelstein et al. 2010), and particularly with our previous paper (Bunker et al.\ 2010). A matched catalog between the Bunker et al.\ (2010), McLure et al.\ (2010) and Bouwens et al.\ (2010a) samples has already been presented in Bunker \& Wilkins (2009).

Our refined HUDF sample, based on a new reduction of the HUDF data, has 6 $Y$-band drop-outs.
In Bunker et al.\ (2010) we presented a list of 7 $Y$-drop candidates within the HUDF field, the brightest four (in $J$-band) of which are reproduced
with the new selection (HUDF-YD1,2,3 \& 4).
Of the 3 other $Y$-drops from Bunker et al.\ (2010), one (YD5) has a discrepant $(Y_{105w}-J_{125w})=0.2$ colour in the new data reduction, much bluer than our selection criteria of $(Y_{105w}-J_{125w})>0.9$. 
The faintest $Y$-drop in Bunker et al.\ (2010), YD7, is marginally too faint ($J=28.65$) in our new reduction of the HUDF images
to enter our new sample. However, applying our new colour selection criteria to the old photometry (where $J=28.44$)
would have resulted in the selection of YD7. 
The remaining one (YD6) is only marginally too blue for the Lyman-break selection in
the newly-reduced data, with $(Y_{f105w}-J_{f125w})=0.89$, very close to the $(Y_{f105w}-J_{f125w})>0.9$ cut. This object has slight ($\sim$\,2\,$\sigma$) detections in the ACS bands, too, and does not meet the selection criterion $(Y_{105w}-J_{125w}) >  0.73\times (J_{125w}-H_{160w})+0.9$, so we did not include it in our list. Moreover, no other group has found or listed this object as a candidate.

Two objects in our new catalog (HUDF.YD8 and HUDF.YD9) were not found in Bunker et al.\ (2010);
our previous study of $Y$-drops in the HUDF used slightly different magnitude and colour cuts ($J_{AB}<28.5$ and $(Y-J)_{AB}>1.0$),
and an older reduction and photometric zeropoints. 
These two objects were slightly too faint in the previous version of our HUDF reductions ($J=28.59$ and $J=28.55$, respectively) and slightly too blue ($(Y_{f105w}-J_{f125w})=0.77$, $0.92$ respectively) to be selected with our original criteria in Bunker et al.\ (2010). 
 The new candidate HUDF.YD8 lies only 1\,arcsec from the $z$-drop zD5 in
Bunker et al.\ (2010), and it is conceivable that both objects might be physically associated and might have similar redshifts at $z\sim 8$. We note that no other group has identified HUDF.YD9 as a candidate.

In Table \ref{tab:common} we show the $Y$-drop galaxy candidates from our HUDF catalog which have been previously reported with their corresponding catalog names from other groups, while in Table \ref{tab:spare} we show all the objects found by these groups with colours or photometric redshifts compatible with being in our $Y$-drop redshift range, but which do not appear in our new catalog. We mark with a $\dagger$ the candidates that would be within our selection window if we adopt the photometry originally presented in the discovery papers, rather than remeasuring this with our new reduction of the HUDF WFC3 imaging and the latest photometric zero-points.

Most of the other HUDF candidates from different groups do not meet our selection criteria both because they are too faint in the $J$-band (class A in the Table \ref{tab:spare}) and because they are too blue, $(Y_{105w} - J_{125w}) < 0.9$ (class B in the Table \ref{tab:spare}).
Only one candidate (z8-SB27 in Yan et al.\ 2010) meets our selection criteria for brightness in the $J$-band and the
$(Y_{105w} - J_{125w})$ colour, but is rejected on the basis of its location in the the $J-H$:$Y-J$ colour:colour plane as a 
likely lower redshift Balmer-break galaxy (see Figure \ref{fig:cc_1}). This galaxy is classified with letter `C' in the table.
We note that  Bouwens' candidate UDFy-37636015 (our YD7) has inconsistent photometry presented in Bouwens et al.\ (2009) and Bouwens et al.\ (2010a) -- adopting the more recent photometric values from Bouwens et al.\ (2010a), YD7 would enter our $Y$-drop selection (Table~\ref{tab:spare}).

In summary, using our latest reduction of the WFC3 images of the HUDF we are able to reproduce 4 of the 7 $Y$-band dropout galaxies we first
reported in Bunker et al.\ (2010); of two additional $Y$-drops in the new analysis, one has been reported elsewhere and
one is a new discovery in the HUDF field. Remeasuring the photometry of $Y$-drop candidates presented elsewhere by
other groups, we find that most would not enter our selection as they are too faint in $J$-band and/or are too blue in $(Y-J)$,
and hence are not as robust candidate $z\approx 8-9$ galaxies as our core sample.

\begin{table}
\begin{tabular}{llllll}
ID & Bo10 & Bu10 & M10 & Y10 & F10 \\
\hline\hline
HUDF-YD2 & UDFy-37796000 & YD2 & 1939y & z8-B117 & 200 \\
HUDF-YD1 & UDFy-42886345 & YD1 & 1765y & z8-B092 & 819 \\
HUDF-YD3 & UDFy-38135539 & YD3 & 1721y & z8-B115 & 125 \\
HUDF-YD4 & - & YD4 & 2487 & - & - \\
HUDF-YD8 & UDFy-43086276 & - & 2841y & z8-B088 & 653 \\
\end{tabular}
\caption{A list of $Y$-drops in the HUDF appearing in the catalogs of all previous analyses. We show in columns the different candidate ID used in this paper, in Bo10 (Bouwens et al.\ 2010a), Bu10 (Bunker et al.\ 2010), M10 (McLure et al.\ 2010), Y10 (Yan et al.\ 2010) and F10 (Finkelstein et al.\ 2010).}
\label{tab:common}
\end{table}
\begin{table}
\begin{tabular}{llllll}
Bunker '10 & M10 & F10 & Y10 & Bo10 & Class \\
\hline\hline
 YD5$\dagger$ & - & - & z8-SD24 & - & AB \\
 YD6$\dagger$ & - & - & - & - & \\
 YD7$\dagger$ &  2079y & 213 & z8-B114 & y37636015 & AB\\
 - & 1107z$^{**}$ & - & - & & B \\
 - & 1422 & 2055 & z8-B041c$\dagger$ & & B \\
 - & - & 800$^*$ & - & & B \\
 - & - & 3022 & - & & AB \\
 - & - & 640 & - & & B \\
 - & - & - & z8-B094$\dagger$ & &  \\
 - & - & - & z8-B087$\dagger$ & & \\
 - & - & - & z8-SB27 & & C \\
 - & - & - & z8-SB30 & & B \\
 - & - & - & z8-SD05 & & AB \\
 - & - & - & z8-SD02 & & AB \\
 - & - & - & z8-SD15 & & AB \\
 - & - & - & z8-SD52 & & AB \\
\end{tabular}

$\dagger$ Object that would meet our selection criteria, assuming original photometry.\\
$^*$ Object 800 appears in versions 1,2 \& 3 of the arXiv:0912.1338 version of Finkelstein et al.\ (2010),
but is absent from version 4 and the Astrophysical Journal paper.\\
$^{**}$ Object also found by Oesch et al. (2009b), named UDFz-44716442 and classified 
as a $z$-drop. We included this object in our table since McLure et al. (2010) quote a 
photometric redshift of 7.60, on the edge of our selection range.

\caption{List of candidates from other studies of the HUDF (see Table \ref{tab:common}) 
quoted by respective authors to be either a $Y$-drop or in our redshift range 
($7.6<z<9.8$, see Section 3) but that we do not recover.
Class A denotes objects too faint in $J$-band ($J_{AB}>28.51$). Class B means that the  colour selection criterion $(Y_{105w}-J_{125w}) > 0.9$.
was not met. Class C is where the $(J-H)$:($Y-J)$ colour:colour rules out selection (i.e.,  $(Y_{105w}-J_{125w}) >  0.73\times (J_{125w}-H_{160w})+0.9$ is not met). Objects marked $\dagger$ would meet our selection criteria if we adopt the original photometry (but which do not make
our selection if we use the photometry from our new reduction of the imaging data).}
\label{tab:spare}
\end{table}

\section{Discussion}

\subsection{The Luminosity Function of $Y$-drops at $z\approx 8-9$}
\label{sec:lumfunc}

\begin{figure}
\centering
\includegraphics[width=9pc]{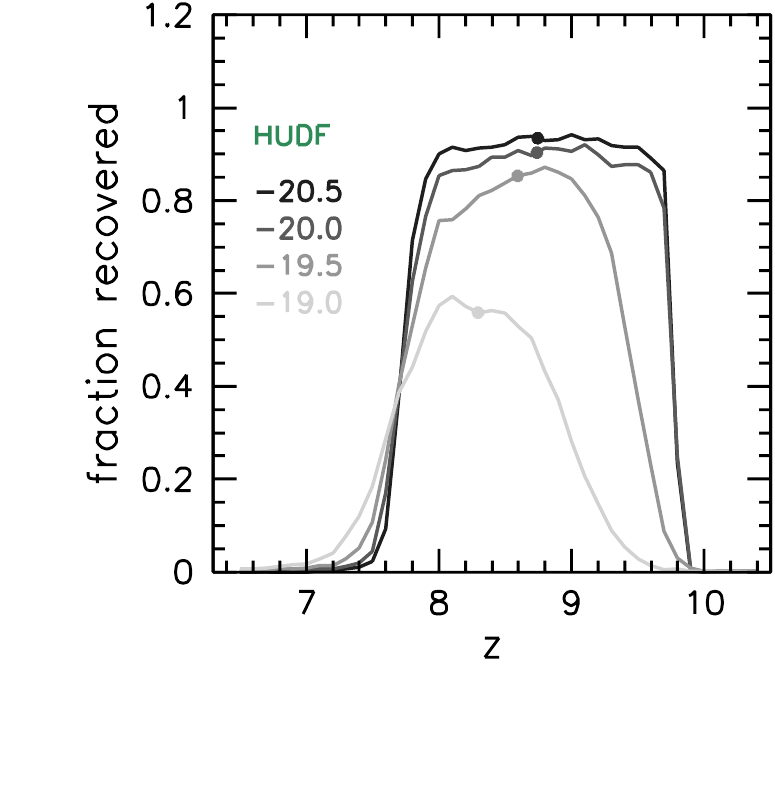}
\includegraphics[width= 9pc]{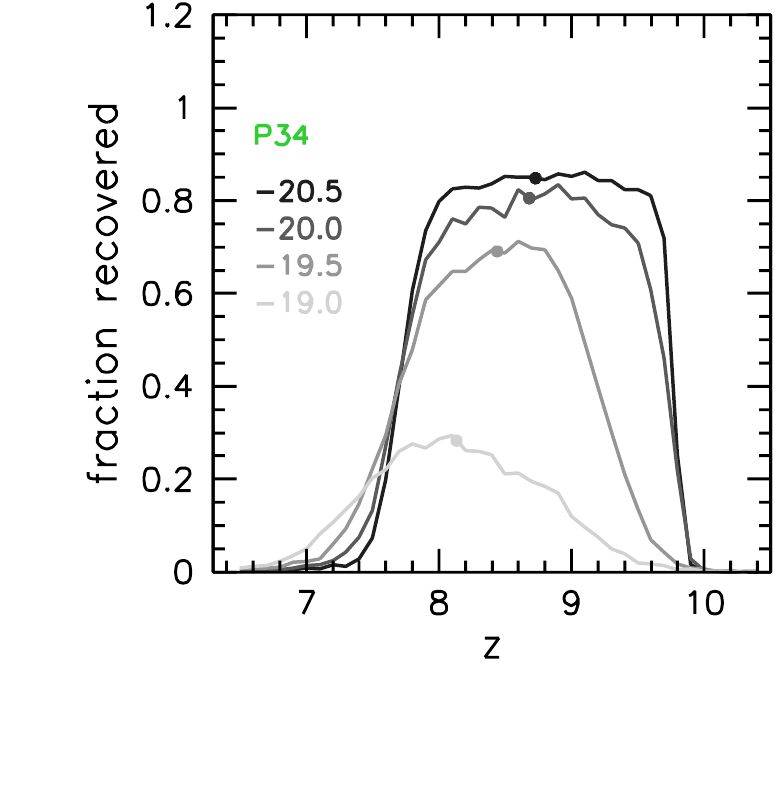}
\centering
\includegraphics[width= 9pc]{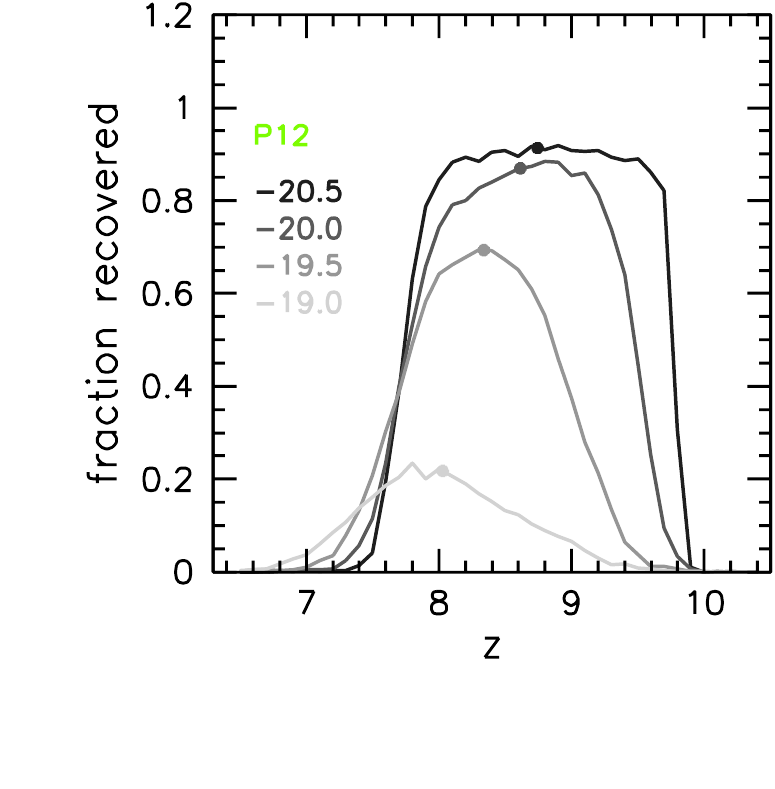}
\includegraphics[width= 9pc]{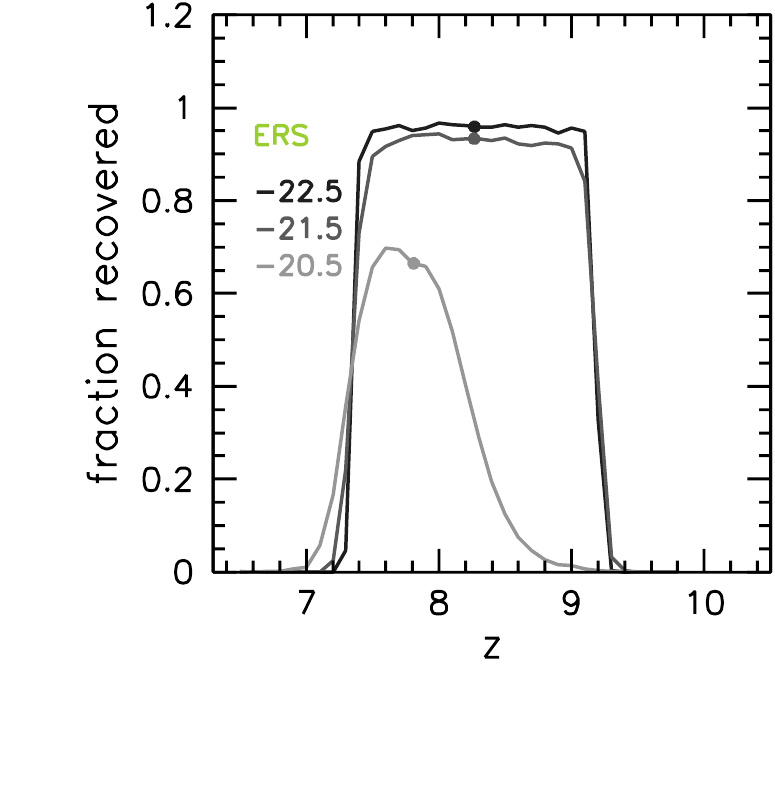}

\caption{The probability of recovering simulated galaxies as a function of redshift and absolute rest-UV magnitude ($M_{1600}$).
We have run simulations on all four of our fields. The mean redshift is denoted by a dot.}
\label{fig:selection}
\end{figure}

From the observed surface density of $Y$-drops, as a function of magnitude, we can recover the luminosity function
of $z\approx 8-9$ galaxies in the rest-frame ultraviolet (observed by the WFC3 near-infrared filters). However, there is not uniform
sensitivity over the redshift range probed by the $Y$-band drop-out technique; at the lower redshifts, the
$Y-J$ colour might be too blue to enter our selection, and at the higher redshift end of our range
the effect of the Lyman-$\alpha$
forest means that an increasingly large fraction of the $J$-band filter is absorbed, so only the most UV-luminous galaxies
will appear in our apparent-magnitude-limited sample. We quantify this effect, and hence constrain the luminosity
function through our observed number counts.
The probability of recovering a high-redshift galaxy as a function of redshift and rest frame UV luminosity  can be found with simulations. To perform these simulations we add into the images a large number of fake galaxies, with properties similar to those of the observed high-redshift population
(i.e. compact with half-light radii $r_{hl}\approx 0\farcs1$, large Lyman-$\alpha$ forest decrement of $D_A\approx 0.99$ and blue rest-frame UV colours). We then run our selection procedure and infer the probability of recovering such galaxies as a function of redshift and magnitude
(see Figure~\ref{fig:selection}).
We adopt the effective volume approach as described in Steidel et al. (1999) and Stanway, Bunker \& McMahon (2003),
such that the probability of recovering a galaxy in our survey depends on the redshift and absolute UV magnitudes, $p(M_{UV},z)$,
and from this the effective survey volume can be calculated ($V_{eff}$). We use a Gaussian distribution of spectral slopes,
with $\langle \beta \rangle = -2.2$ and $\sigma(\beta)=0.5$, reflecting the generally blue spectral slopes observed in Lyman-break
galaxies at $z\ge6$ (Stanway, McMahon \& Bunker 2005; Bouwens et al.\ 2010b; Wilkins et al.\ 2010c; Bunker et al.\ 2010).
In Table~\ref{tab:objects} we have presented our list of candidate $Y$-drops, with colours consistent with being high redshift.
These are good targets for spectroscopy, but in calculating the luminosity function we wish to restrict the sample to only
the most reliable sources (to minimize biases through contamination by photometric scatter).
In determining the luminosity function, we do not consider the P12 $Y$-drops, where the shallow $H$-band means we
do not have secure $H$-band magnitudes. For the other fields, we include only those galaxies from Table~\ref{tab:objects} detected 
at $\ge 7\,\sigma$ in $J$-band which are not flagged as marginal;
the only galaxies at $>7\,\sigma$ in $J$-band 
not included in the luminosity function fits are ERS.YD2 
and ERS.YD7.

We can then determine the best-fit luminosity function (Figure \ref{fig:lf_Y}); we assume a Schechter (1976) functional form,
where the number density of galaxies between $L$ and $L+\delta L$ is:
\[ \phi(L)dL = \phi^* e^{-x} x^\alpha
\]
where $x=L/L^*$. The Schechter function is
parameterized by a faint end slope ($\alpha$), a characteristic number density at the knee of the luminosity function ($\phi^*$)
and the characteristic luminosity, $L^*$, corresponding to the absolute magnitude in the rest-frame UV ($M^*_{UV}$, determined around 1600\,\AA ).
Unfortunately we still do not have enough faint galaxies to constrain the faint end slope of this function, so we adopt three different values for the faint end slope, $\alpha =-1.5$, $-1.7$, $-1.9$, bracketting the value of $\alpha=-1.73$ derived by Bouwens et al.\ (2006) for the $i'$-drops
at $z = 6$ and for the $z=3$ $U$-drops (Reddy \& Steidel 2009). We fit for the free parameters $\phi^*$ and $M^*_{UV}$, and these
are presented in Table~\ref{tab:uvld}.

\begin{table*}
\begin{tabular}{cccccc}
$\alpha$ & $M_{1600}^{*}$ [AB mag] & $\phi^{*}$ [Mpc$^{-3}$] & \multicolumn{3}{|c|}{$\rho_{1600}\,[10^{25}\,{\rm erg\, s^{-1}\, Mpc^{-3}\, Hz^{-1}}]$ $
(\dot{\rho_{*}}\,[{\rm M_{\odot}\,yr^{-1}\,Mpc^{-3}}])$} \\
 & & & $M_{1600}<-18.5$ $({\rm SFR}>1.5\,{\rm M_{\odot}\,yr^{-1}})$ & $<-13$ $(>0.01\,{\rm M_{\odot}\,yr^{-1}})$ & $<-8$ $(>10^{-4}\,{\rm M_{\odot}\,yr^{-1}})$ \\
\hline\hline
$-$1.5 & $-$19.34 & 0.00117 &  1.65 ( 0.0022 ) & 4.61 ( 0.0060 ) & 4.88 ( 0.0064 ) \\
$-$1.7 & $-$19.5 & 0.00093 &  1.71 ( 0.0022 ) & 6.22 ( 0.0081 ) & 7.27 ( 0.0095 ) \\
$-$1.9 & $-$19.66 & 0.00070 & 1.73 ( 0.0023 ) & 9.05 ( 0.0119 ) & 13.46 ( 0.0176 ) \\
\end{tabular}
\caption{The best fit values of $M^{*}_{1600}$ and $\phi^{*}$ for s Schechter function assuming fixed $\alpha\in\{-1.5,-1.7,-1.9\}$ together with the UV luminosity densities (and star formation rate densities in parentheses) determined by integrating the luminosity function down to various limiting absolute magnitudes.}
\label{tab:uvld}
\end{table*}

\subsection{Evolution of the Luminosity Function with Redshift}

\begin{figure}
\centering
\includegraphics[width=20pc]{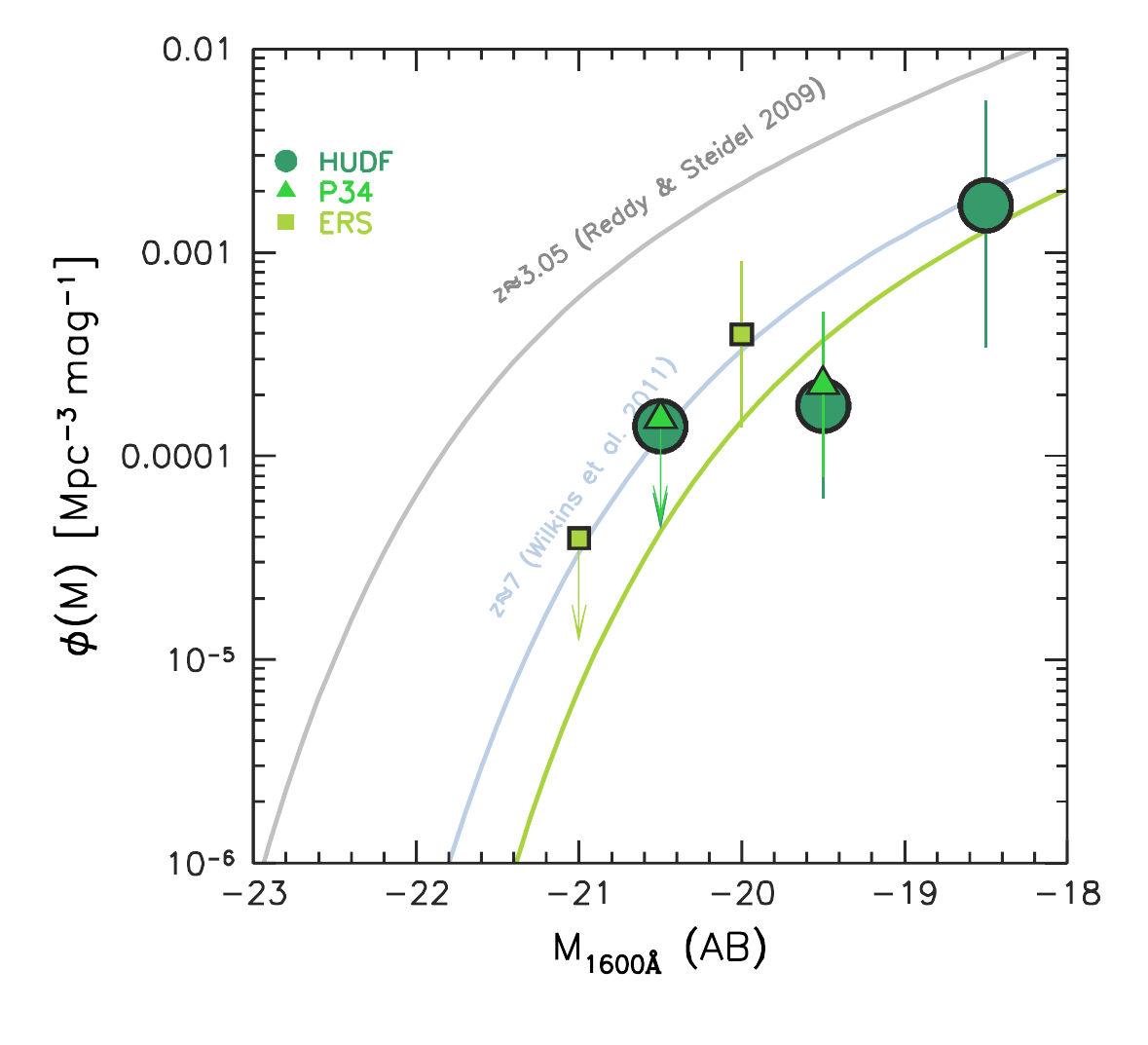}
\caption{The $z\approx 8-9$ rest-frame UV ($1600{\rm\AA}$) luminosity function derived from HUDF, UDF-P34 and ERS WFC3 fields (circles, triangles and squares, respectively) together with contemporary and lower-redshift comparisons. Solid lines denote the luminosity function at $\langle z\rangle=3.05$ (Reddy \& Steidel 2009) and $\langle z\rangle \approx=7.0$ (Wilkins et al. 2010b). 
}
\label{fig:lf_Y}
\end{figure}

\begin{figure}
\centering
\includegraphics[width=20pc]{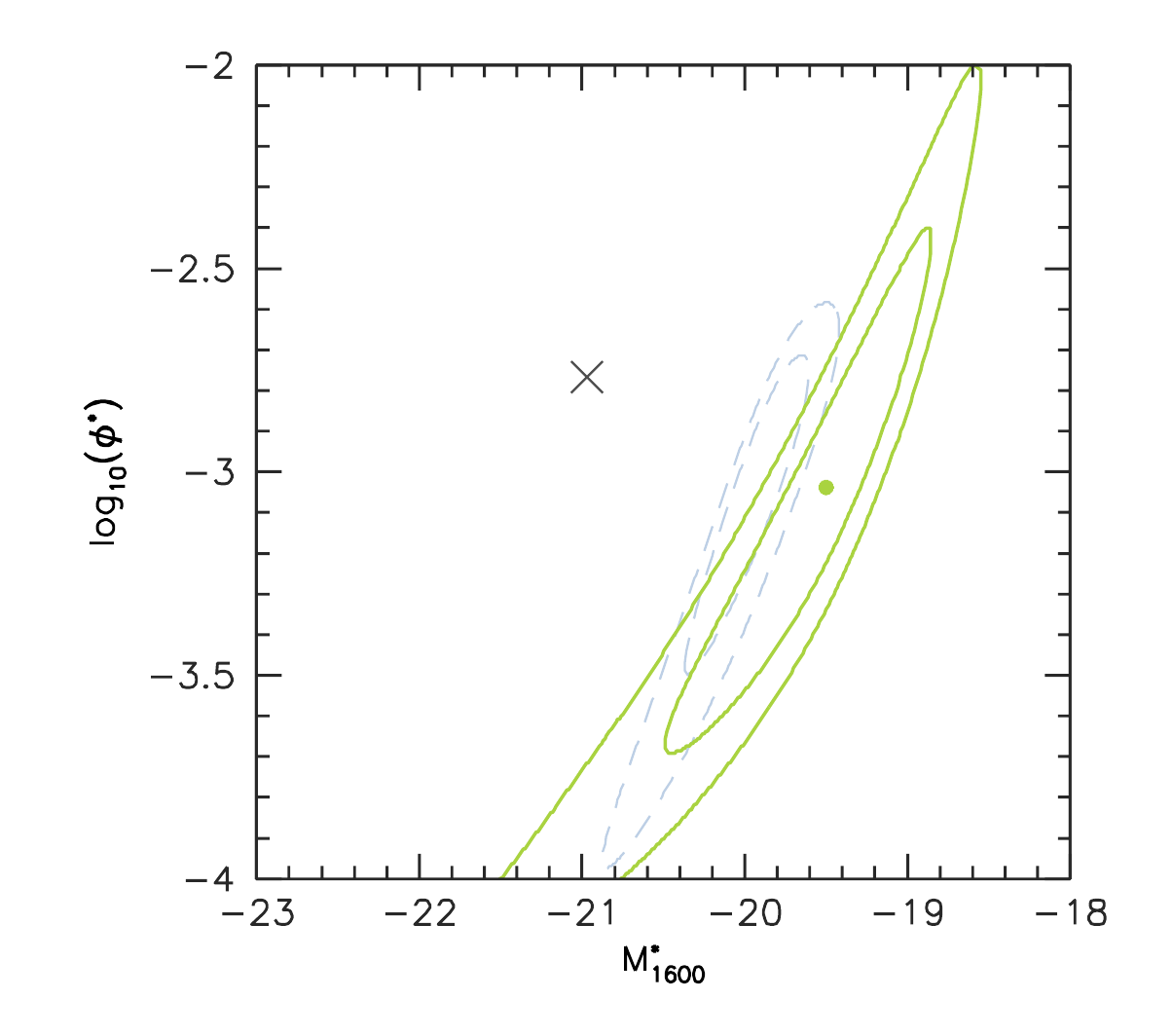}
\caption{The significance contours for the reduced-$\chi^2$ fits of the Schechter UV luminosity function for the $Y$-drops
(solid contours on right, signifying reduced-$\chi^2=1$ and $2$ for the inner and outer). A faint end slope of $\alpha=-1.7$
has been assumed, and the dot is the formal best fit, with $M^*_{1600}=-19.5$ (AB). The dashed contours (to the left of the $z=8$
contours) denote the $z=7$ luminosity function derived by Wilkins et al.\ (2010b) from the $z$-drops in the same WFC\,3 fields
as analysed here. The cross (on the left) is the $z=3$ luminosity function for Lyman-break galaxies (Reddy \& Steidel 2009).
Evolution predominantly in $M^*$  is most consistent with the observational data.}
\label{fig:reducedChi2}
\end{figure}

We now compare our measured best-fit luminosity function parameters (Table 6) with previous estimates from the $z\approx 8-9$ galaxies in the HUDF
alone. We also plot the $z=7$ UV luminosity function from the $z$-drops of Wilkins et al.\ (2010b), and note that other LFs based on smaller datasets have derived similar parameters (e.g.\ McLure et al.\ 2010; Oesch et al.\ 2010; Ouchi et al.\ 2010). Based on 5 $Y$-drops, Bouwens et al.\ (2010a) estimated $M^*_{UV}=-19.45$, assuming no evolution in $\phi^*$ and $\alpha$ from $z\approx 6$ (fixing $\phi^*=0.0011\,{\rm Mpc}^{-3}$ and $\alpha=-1.74$). This is consistent with our determination of $\phi^*=0.00093\,{\rm Mpc}^{-3}$ and $M^*_{UV}=-19.5$ (where we have fixed $\alpha=-1.7$ but fit both $\phi^*$ and $M^*_{UV}$). We note that our measured charateristic number density
is within $\approx 20$ per cent of the Bouwens et al.\ assumption, with $M^*_{UV}$ nearly the same as the Bouwens et al.\ fit.
McLure et al.\ (2010) suggest that the main luminosity function evolution from $z\approx 6$ is in $\phi^*$, with $M^*_{UV}$ broadly unchanged
at $M_{UV}=-20$, and $\phi^*_{z=6}\approx 5\times \phi^*_{z=8}$. However, this appears to be marginally inconsistent with our number counts of $Y$-drops at the bright end --
this parameter space is unavailable using the HUDF alone, but the larger volume we have in our current study enables us to fit both
$\phi^*$ and $M^*_{UV}$ at $z\approx 8$ -- each of the 6 independent points in the luminosity function (Figure~\ref{fig:lf_Y}) has $\approx 3$ galaxies
in it (rather than 2 bins of $2-3$ galaxies previously). The McLure et al.\ (2010) pure-density-evolution scenario lies on the $1\,\sigma$ contour of our $\phi^*$ vs.\ $L^*$ reduced-$\chi^2$ plot (Figure \ref{fig:reducedChi2}). Our results are entirely consistent with an evolution mainly in $M^*_{UV}$ since  $z=7$ (and indeed since $z=3$), with only a modest change in $\phi^*$ (consistent with no change in $\phi^*$).

\begin{figure}
\centering
\includegraphics[width=20pc]{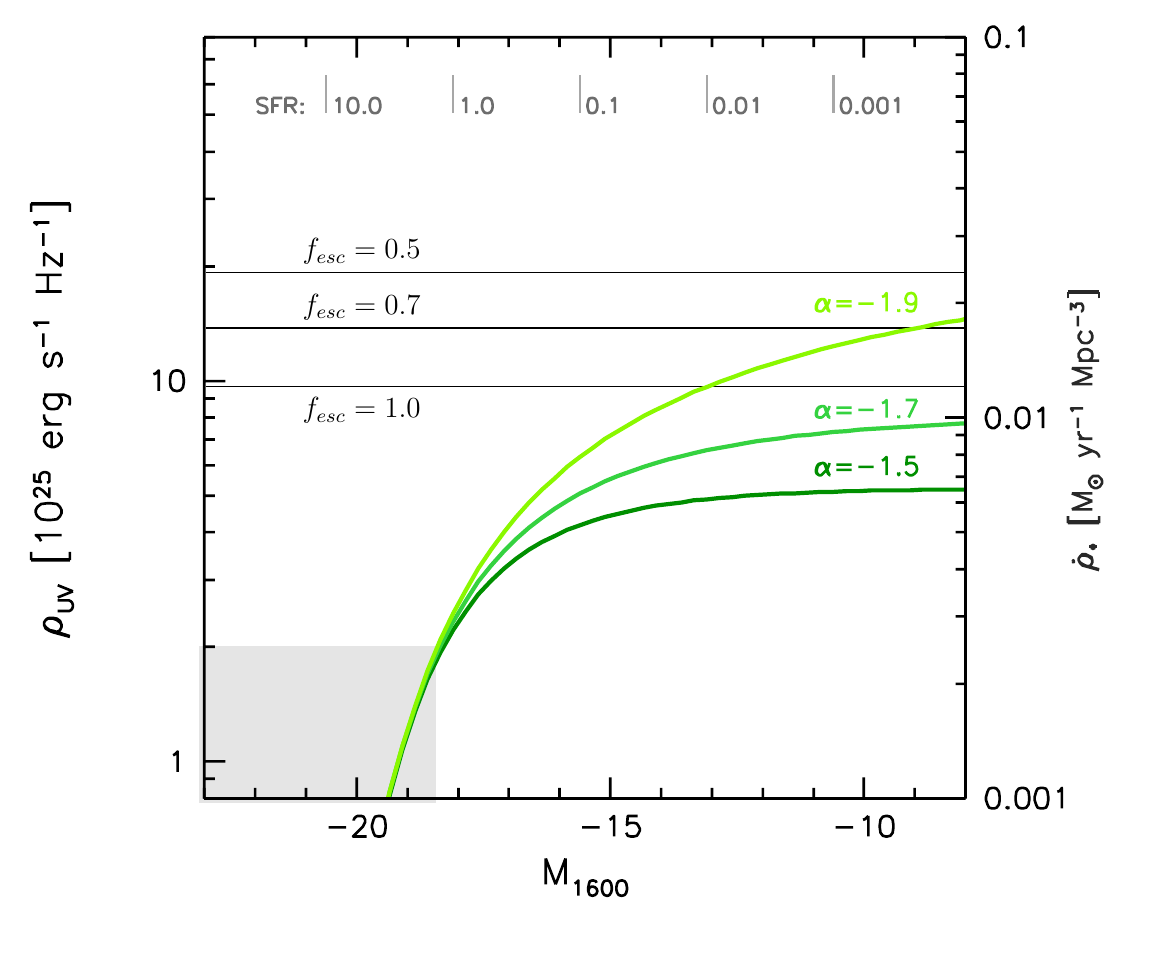}
\caption{The UV luminosity density (left axis) and star forma- 
tion rate density (right axis) as a function of the rest-UV ($M_{1600}$) absolute magnitude 
down to which the luminosity function is integrated. We show our best 
fit luminosity functions at $z\approx 8-9$ assuming $\alpha$ = {$-$1.5, $-$1.7, $-$1.9}. The shaded grey box denotes the observed region with the remainder inferred from extrapolation of the luminosity function. The horizontal lines show the UV luminosity density required
to reionize the Universe at this redshfit, assuning a clumping factor of $C=5$, and an escape fraction of $f_{\mathrm{esc}}=0.5$ (top line), $0.7$ (middle) and  $1$ (bottom).}
\label{fig:sfrd_Y}
\end{figure}

The integration of the luminosity function gives us the ultraviolet luminosity density, which is important for our purposes, since it is directly connected with the star formation rate density. Figure \ref{fig:sfrd_Y} shows the UV luminosity density as a function of the magnitude down to which the luminosity function is integrated for each of the three best fits (one for each value of $\alpha$ adopted).

\subsection{The Star Formation Rate Density at $z\approx 8$}

We can use the observed $H$-band magnitudes of objects in the $Y$-drop sample to estimate their star formation rate from the rest-frame UV luminosity density around $\lambda_{rest}=1600$\,\AA . 
In the absence of dust
obscuration, the relation between the flux density in the rest-UV
around this wavelength and the star formation rate (${\rm SFR}$
in $M_{\odot}\,{\rm yr}^{-1}$) is given by $L_{\rm UV}=8\times 10^{27}
{\rm SFR}\,{\rm ergs\,s^{-1}\,Hz^{-1}}$ from Madau, Pozzetti \&
Dickinson (1998) for a Salpeter (1955) stellar initial mass function
(IMF) with $0.1\,M_{\odot}<M_{*}<125\,M_{\odot}$. This is comparable
to the relation derived from the models of Leitherer \& Heckman (1995)
and Kennicutt (1998).  However, if a Scalo (1986) IMF is used, the
inferred star formation rates will be a factor of $\approx 2.5$ higher
for a similar mass range.
In the absence of a spectroscopic redshift, we assume that these lie at the predicted average redshift for galaxies obeying our colour cuts (Figure~\ref{fig:selection}). For the luminosity functions considered, the predicted mean redshift is around $\langle z\rangle=8.6$ for a spectral slope $\beta \approx -2$
and $M_{UV}=-20.5$. The $H$-band probes the rest-UV above Lyman-$\alpha$, and is unaffected by the forest absorption.
Our $6\,\sigma$ limit for the HUDF is $H_{AB}=28.5$, equivalent to an absolute magnitude of $M_{1600\rm \AA }=-18.6$ at $z=8.6$, corresponding to an inferred star formation of $1.5\,M_{\odot}\,{\mathrm{yr}}^{-1}$. This is equivalent to $0.1\,L^*_{z=3}$ for $M^*_{UV}=-20.97$  at $z=3$ (Reddy \& Steidel 2009) .

With our measured Schechter luminosity function parameters ($\phi^*=0.00093$\,Mpc$^{-3}$, $M^*=-19.5$ assuming $\alpha=-1.7$ as at $z=3$ and $z=6$), the total star formation rate density is $0.0022\,M_{\odot}\,{\rm yr^{-1}\,Mpc^{-3}}$ integrating down to our luminosity limit of $M^{UV}_{\rm 1600\,\AA }\approx-18.5$\,mag (AB). This should be regarded as a robust {\em lower limit} on the star formation rate density, as dust obscuration may affect the rest-frame UV continuum, and also galaxies fainter than our selection limit will also contribute to the integrated UV light density. If instead we integrate down to $M^{UV}_{\rm 1600\,\AA }=-13$ (corresponding to $0.01\,M_{\odot}\,{\rm yr^{-1}}$) and the total star formation rate density is $0.0081\,M_{\odot}\,{\mathrm{yr}}^{-1}\,{\mathrm{Mpc}}^{-3}$. These star formation rate densities are a factor of $\sim 10$ {\em lower} than at $z\sim 3-4$, and even a factor of $\approx 3-5$ below that at $z\approx 6$ (Bunker et al.\ 2004; Bouwens et al.\ 2006).

\subsection{Implications for Reionization}

The ionizing UV photons produced by the most massive (OB) stars might be critical in reionization and keeping the Universe ionized at $z\approx 6-11$. However, work at $z\approx 6$ has shown that under standard assumptions of the IMF, escape fraction and clumping of the gas, the observed population of Lyman break galaxies produce insufficient flux down to $AB\approx 28.5$\,mag (Bunker et al.\ 2004), and the ``photon drought" is even more severe at $z\approx 7$ (Wilkins et al.\ 2010b). We now compare our measured UV luminosity density at $z\approx 8-9$ (quoted above as a corresponding star formation rate density) with that required to ionize the Universe at this redshift.
Madau, Haardt \& Rees (1999) give the density of star formation
required for reionization (assuming the same Salpeter IMF as used in this paper):
\[
 {\dot{\rho}}_{\mathrm{SFR}}\approx \frac{0.012\,M_{\odot}\,{\mathrm{yr}}^{-1}\,{\mathrm{Mpc}}^{
-3}}{f_{\mathrm{esc}}}\,\left( \frac{1+z}{1+8.6}\right) ^{3}\,\left( \frac{\Omega_{b}\,h^
2_{70}}{0.0462}\right) ^{2}\,\left( \frac{C}{5}\right)
 \]
We have updated equation 27 of Madau, Haardt \& Rees (1999) for a more recent concordance cosmology estimate of the baryon
density from Larson et al.\ (2010), $\Omega_b\,h_{100}^2=0.022622$. The reionization requirement at $z\approx 8.6$ is
a factor of 2.5 times higher than that at $z\approx 6$, as the number of photons needed rises as $(1+z)^3$.

In the above equation, $C$ is the clumping factor of neutral
hydrogen, $C=\left< \rho^{2}_{\mathrm{HI}}\right> \left< \rho_{\mathrm{HI}}\right> ^{-2}$. Early simulations suggested $C\approx 30$ (Gnedin \&
Ostriker 1997), but more recent work including the effects of reheating implies a lower
concentration factor of  $C\approx 5$ (Pawlik et al.\ 2009). 
The escape fraction of ionizing photons ($f_{\mathrm{esc}}$) for
high-redshift galaxies is highly uncertain (e.g., Steidel, Pettini \&
Adelberger 2001, Shapley et al.\ 2006), and it is possible
that the escape fraction of ionising photons may be linked to the escape fraction of Lyman-$\alpha$ photons (Stark et al.\ 2010),
which may mean that high escape fractions could be tested through future line emission line searches with spectroscopy and narrow-band imaging.
Even if we take the upper limit of $f_{\mathrm{esc}}=1$ (no absorption by
H{\scriptsize~I}) and a very low clumping factor, the required total star formation rate density for reionization
is $0.012\,M_{\odot}\,{\mathrm{yr}}^{-1}\,{\mathrm{Mpc}}^{
-3}$. This is  a factor of $\sim 5$ higher than our measured star formation density at
$z\approx 8-9$ from $Y$-drop galaxies brighter than $M_{UV}=-18.5$ (our approximate limit). As shown in Table~\ref{tab:uvld} and Figure~\ref{fig:sfrd_Y}, 
the required UV luminosity density can only just be achieved (if $f_{\mathrm{esc}}=1$) by integrating down to $M_{UV}=-13$ (i.e., extrapolating the Schechter function to $\approx 100$ times fainter than our observed limit) and then only for a steeper faint end slope of $\alpha=-1.9$
rather than $\alpha=-1.7$. Adopting a less unrealistic value of $f_{\mathrm{esc}}=0.7$ (which is still high compared with observed values at lower redshift) the required total star formation rate density for reionization would be $0.017\,M_{\odot}\,{\mathrm{yr}}^{-1}\,{\mathrm{Mpc}}^{
-3}$, then the $Y$-drop population can only provide sufficient ionizing photons if the faint end slope is very steep ($\alpha\le-1.9$) and the Schechter function is integrated down below $M_{UV}=-8$ (corresponding to a star formation rate of only $10^{-4}\,M_{\odot}\,{\mathrm{yr}}^{-1}$). We note that recent theoretical papers indicate that the reionization process itself may have been ``photon-starved" (e.g., Bolton \& Haehnelt 2007), consistent with the extrapolation of our observational constraints.

However, the assumption of a solar metallicity Salpeter IMF may be flawed: the colours of $z\sim 6$ $i'$-band drop-outs are very blue (Stanway, McMahon \& Bunker 2005), with $\beta<-2$, and the recent WFC3 $J$- and $H$-band images show that the $z\approx 7$ $z'$-drops also have blue colours on average (Bunker et al.\ 2010; Bouwens et al.\ 2010b; Wilkins et al.\ 2010c). 
Continuous star formation with a Salpeter IMF produces a UV spectral slope of $\beta\approx -2$  if there is no dust reddening. The fact
that we observe even more blue slopes than this ($\beta<-2$) could be explained through low metallicity,
 or a top-heavy IMF, which can produce between 3 and 10 times as many ionizing photons for the same 
 1600\,\AA\ UV luminosity (Schaerer 2003 -- see also Stiavelli, Fall \& Panagia 2004). Alternatively, we may be seeing galaxies at the onset of star formation, or with a rising star formation rate (Verma et al.\ 2007), which would also lead us to underestimate the true star formation rate from the rest-UV luminosity. We explore the implications of the blue UV spectral slopes in $z\ge 6$ galaxies in a forthcoming paper (Wilkins et al.\ 2010c).

\section{Conclusions}

In this paper we have presented a search for galaxies at $7.5<z<10$ using the latest {\em HST} WFC3 near-infrared data,
based on the Lyman-break technique. Searching for galaxies which have large $(Y-J)$ colours ($Y$-drops) on account of
the Lyman-alpha forest absorption, and with $(J-H)$ colours inconsistent with being low-redshift contaminants,
we identify $\approx 20$ candidates at redshift $z \approx 8 - 9$ over an area of $\approx 50$ square arcminutes.
Our deepest field (the HUDF, covering 4.2\,arcmin$^2$) reaches $J_{AB}=28.5$ at $6\,\sigma$, while the
wide-area ERS data (comprising 10 WFC3 pointings covering 37\,arcmin$^2$) reaches $J_{AB}=27.2$.
The surface densities of candidates as a function of limiting magnitude appear broadly consistent between our
4 fields, although these all lie within 10\,arcmin.
Previous searches for $Y$-drops with WFC3 have focussed only on the HUDF, and our larger survey has trebelled
the number of robust $Y$-drop candidates, as well as providing a number of brighter $Y$-drops
(with $J_{AB}\approx 27.0$ rather than $J_{AB}>28.0$ as in the HUDF). These brighter sources
may be more amenable to spectroscopic follow-up. 

For the first time, we have a sufficient number of $z\approx 8-9$ galaxies
to fit $\phi^*$ and $M^*$ assuming a Schechter luminosity function (previous estimates had to fix one of these
parameters). We confirm that there is large evolution from $z=3$, particularly in the bright end of the luminosity
function, in the sense that there are far fewer UV-bright galaxies at $z\approx 8-9$ than in the more
recent past. There is also evidence for evolution from $z=6-7$ to $z=8-9$, with this being
consistent with most of the change occurring in $M^*$ rather than $\phi^*$, with $M^*$ being fainter
at higher redshift. We are unable to obtain a good constraint on the faint-end slope, $\alpha$, which will
potentially require deeper data over a wider field (as might be provided by NIRCAM on the {\em James
Webb Space Telescope}). The candidate $z\approx 8-9$ galaxies we detect have insufficient ionizing
flux to reionize the Universe, and it is probable that galaxies below our detection limit provide a significant
UV contribution. However, adopting a similiar faint-end slope to that determined at $z=3-6$ ($\alpha=-1.7$)
and a Salpeter IMF,
then the ionizing photon budget still falls short if $f_{\mathrm{esc}}<0.5$, even integrating down to $M_{UV}=-8$.
A steeper faint end slope and a low-metallicity population (or a top-heavy IMF) might still provide sufficient
photons for star-forming galaxies to reionize the Universe, but confirmation of this might have to await
the {\em James Webb Space Telescope}.

\subsection*{Acknowledgements}
Based on observations made with the NASA/ESA Hubble Space Telescope,
obtained from the Data Archive at the Space Telescope Science Institute, which is operated by the Association
of Universities for Research in Astronomy, Inc., under NASA contract
NAS 5-26555. These observations are associated with programme \#GO-11563 and \#GO/DD-11359. We are grateful to
to Garth Illingworth and his team, and the  WFC\,3 Science Oversight Committee for making their Early Release Science and
Hubble Ultra Deep Field observations public. We thank Richard Ellis for useful comments on this paper.
MJJ acknowledges the support of a RCUK
fellowship. 
SL and JC are supported by the Marie Curie Initial Training Network ELIXIR of the European Commission under
contract PITN-GA-2008-214227. We thank the anonymous referee for helpful comments which have improved this paper.

\bsp

\end{document}